\documentclass[structabstract]{aa}
\pdfoutput=1
\usepackage{txfonts}
\usepackage{pictex}
\usepackage{latexsym}
\usepackage{graphics}
\usepackage{afterpage}
\usepackage{graphicx}

\begin{document}

\title
{
A large sample of Kohonen-selected SDSS quasars with weak emission lines: 
selection effects and statistical properties\thanks{The full catalogue is only available in 
electronic form at the CDS via anonymous ftp to cdsarc.u-strasbg.fr}
}

\author{H. Meusinger   \inst{1}
        \and
        N. Balafkan      \inst{2}
       }
  
    \institute{
     \inst{1} Th\"uringer Landessternwarte Tautenburg, Sternwarte 5, 07778 Tautenburg, Germany,
          e-mail: meus@tls-tautenburg.de \\
     \inst{2} Universit\"at Leipzig, Faculty of Physics and Earth Sciences, Linn\`estr. 5, 04103 Leipzig, Germany,
          e-mail: n.balafkan@studserv.uni-leipzig.de \\
        }
\date{Received / Accepted }

\abstract {}
   {
   A tiny fraction of the quasar population shows remarkably weak emission lines. Several hypotheses have been developed,
   but the weak line quasar (WLQ) phenomenon still remains puzzling. The aim of this study was to create a sizeable sample
   of WLQs and WLQ-like objects and to evaluate various properties of this sample.
   }
   {
   We performed a search for WLQs in the spectroscopic data from the Sloan Digital Sky Survey Data Release 7
   based on Kohonen self-organising maps for nearly $10^5$ quasar spectra. The final sample consists of 365 quasars
   in the redshift range $z = 0.6 - 4.2$ ($\overline{z} = 1.50 \pm 0.45$) 
   and includes in particular a subsample of 46 WLQs with equivalent widths $W_\ion{Mg}{ii} < 11$\AA\ 
   and $W_\ion{C}{iv} < 4.8$\AA.
   We compared the luminosities, black hole masses, Eddington ratios, accretion rates,
   variability, spectral slopes, and radio properties of the WLQs with those of control samples of ordinary quasars.
   Particular attention was paid to selection effects.
   }
   {
   The WLQs have, on average, significantly higher luminosities, Eddington ratios, and accretion rates.
   About half of the excess comes from a selection bias, but an intrinsic excess remains probably caused 
   primarily by higher accretion rates. The spectral energy distribution shows a bluer continuum
   at rest-frame wavelengths $\ga$ 1500 \AA. The variability in the optical and UV is relatively low, 
   even taking the variability-luminosity anti-correlation  into account. The percentage of radio detected 
   quasars and of core-dominant radio sources is significantly higher than for the control sample, whereas 
   the mean radio-loudness is lower.
   }
   {
   The properties of our WLQ sample can be consistently understood assuming that it consists of a mix of quasars at
   the beginning of a stage of increased accretion activity and of beamed radio-quiet quasars. The higher luminosities and
   Eddington ratios in combination with a bluer spectral energy distribution
   can be explained by hotter continua, i.e. higher accretion rates.
   If quasar activity consists of subphases with different accretion rates, a change towards a higher rate is
   probably accompanied by an only slow development of the broad line region. The composite WLQ 
   spectrum can be reasonably matched by the ordinary quasar composite where the continuum has been replaced by that 
   of a hotter disk. A similar effect can be achieved by an additional power-law component in relativistically
   boosted radio-quiet quasars, which may explain the high percentage of radio quasars.
   }      

\titlerunning{SDSS quasars with weak emission lines}
\authorrunning{H. Meusinger \& N. Balafkan}

\maketitle

%
%
\section{Introduction}
%
%

\begin{figure*}[bhpt]
\begin{tabbing}
\includegraphics[viewport=50 40 500 800,angle=270,width=6.6cm,clip]{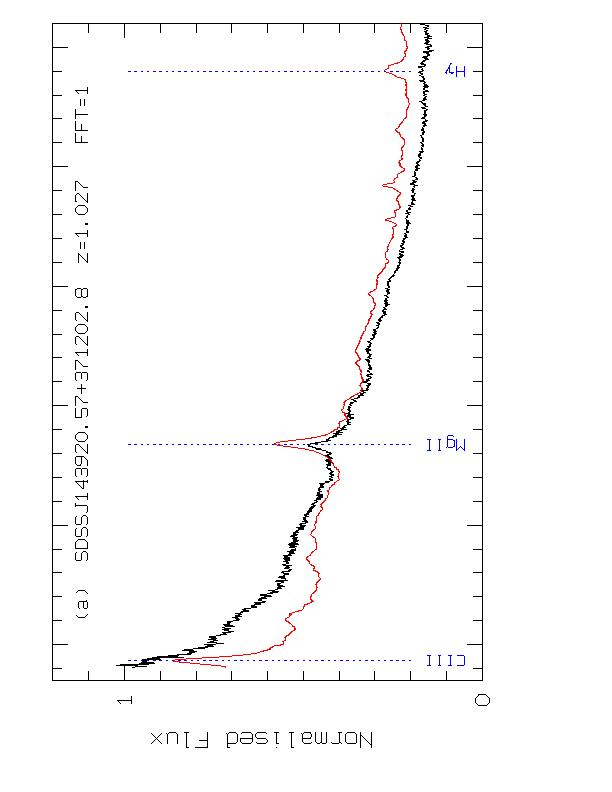}\hfill
\includegraphics[viewport=50 110 500 800,angle=270,width=6.0cm,clip]{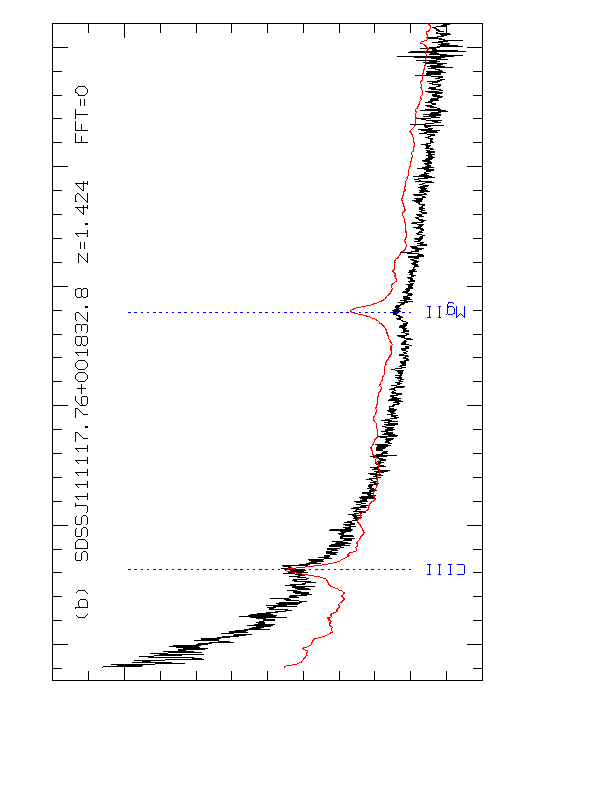}\hfill
\includegraphics[viewport=50 110 500 800,angle=270,width=6.0cm,clip]{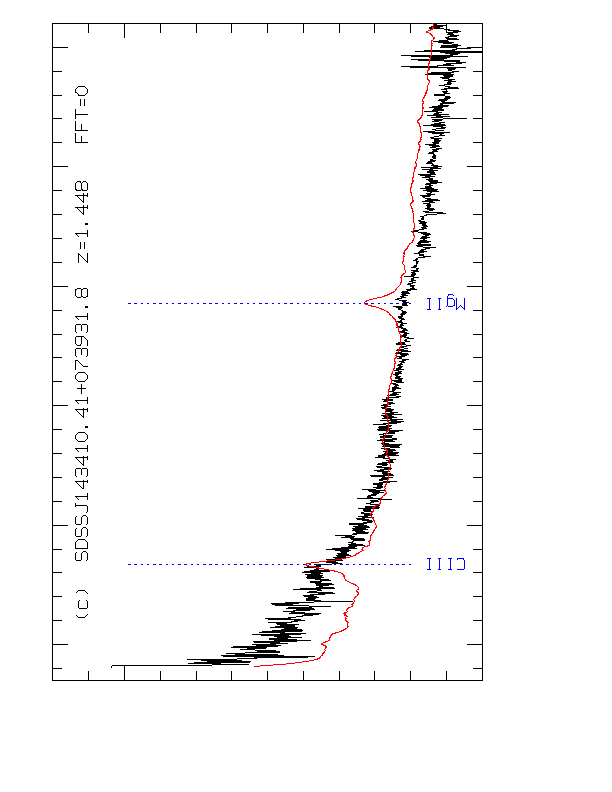}\hfill \\
\includegraphics[viewport=50 40 560 800,angle=270,width=6.6cm,clip]{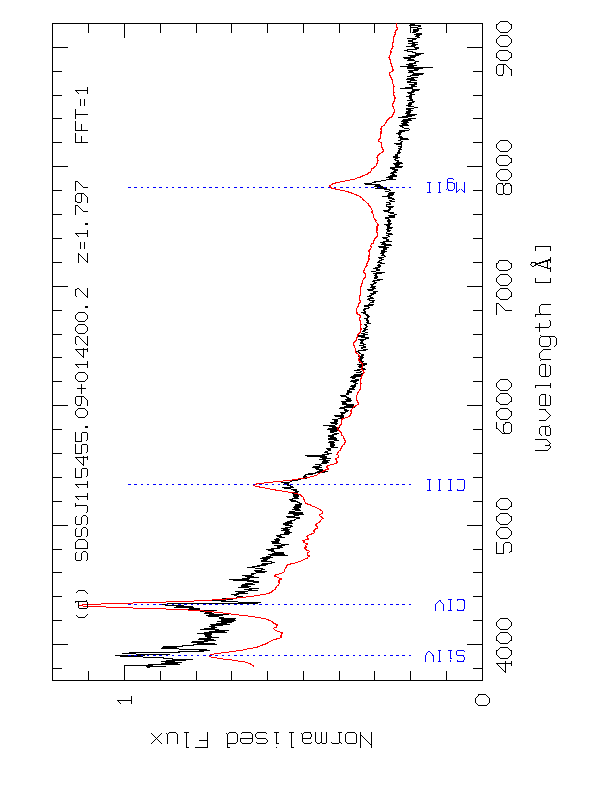}\hfill
\includegraphics[viewport=50 110 560 800,angle=270,width=6.0cm,clip]{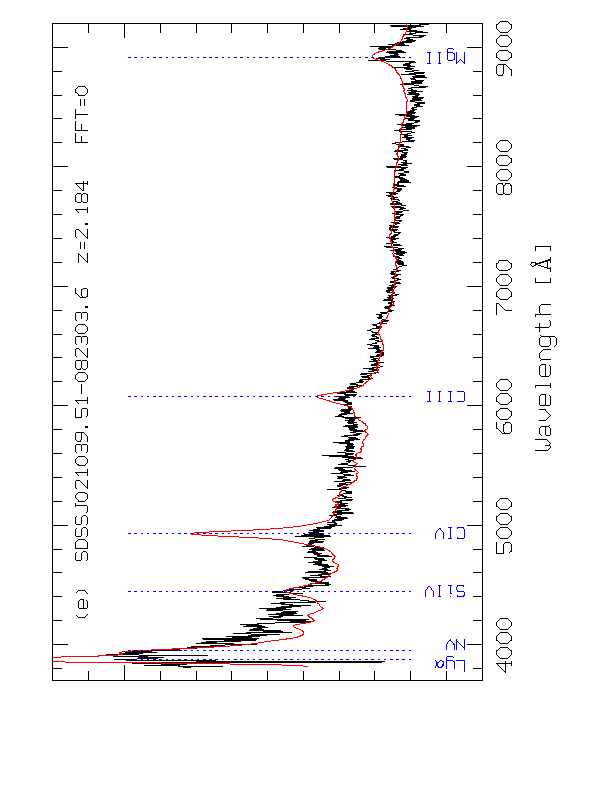}\hfill
\includegraphics[viewport=50 110 560 800,angle=270,width=6.0cm,clip]{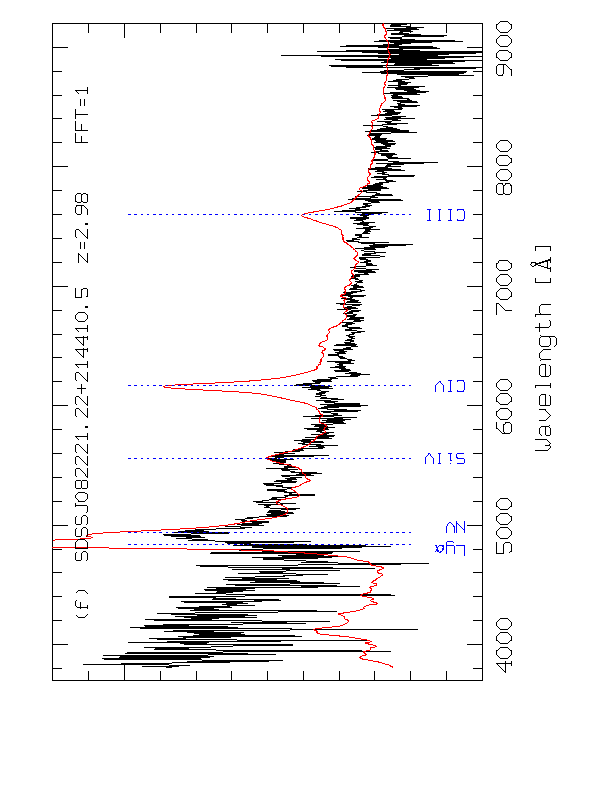}\hfill
\end{tabbing}
\caption{
Six examples of quasars with weak emission lines from the WLQ sample.
For comparison the quasar composite spectrum from 
Vanden Berk et al. (\cite{VandenBerk2001}; smooth red curve)
is over-plotted, shifted to the redshift 
of the quasar and normalised to the total flux density between 3900 \AA\ and 9000 \AA\ (observer frame).
At the top of each panel, the SDSS name, the redshift $z$, and the parameter FFT = {\sc first\_fr\_type} 
(0 - not detected by FIRST; 1 - core-dominated radio source) from the Shen catalogue are given.
The dotted vertical lines indicate the usually strong emission lines.
}
\label{fig:examples}
\end{figure*}

Broad emission lines (BELs) are a defining characteristic of type 1 active galactic nuclei (AGN). 
The weakness or even absence of BELs is the most remarkable feature
of a class of high-luminosity AGNs called weak-line quasars (WLQs).
The first discovered WLQ was the radio-quiet quasar PG 1407+265 at redshift $z = 0.94$
(McDowell et al. \cite{Mcdowell1995}) with undetectably weak H$\mathrm{\beta}$ and UV 
BELs although the continuum properties are similar to those of normal radio-quiet quasars.
Fan et al. (\cite{Fan1999}) discovered the first high-$z$ WLQ, SDSS J153259.96+003944.1 
($z = 4.62$) and suggested that it is either the most distant known BL Lac object 
with very weak radio emission or a new type of unbeamed quasars whose broad emission line 
region (BLR) is very weak or absent.
Based on the multi-colour selection of the Sloan Digital Sky Survey (SDSS;
York et al. \cite{York2000}), about one hundred high-$z$ WLQs have been found with Ly$\alpha$-\ion{N}{v}
rest-frame equivalent width $< 15$ \AA\ (Diamond-Stanic et al. \cite{Diamond2009}; 
Shemmer et al. \cite{Shemmer2010}; Wu et al. \cite{Wu2012}).

Low values of the equivalent widths of the BELs can be the result of abnormally low
line fluxes or of an unusually strong continuum. Relativistic beaming provides an
example for dilution of the line strength by a boosted continuum. However, such an 
interpretation  of the WLQ phenomenon is widely considered unlikely because many properties of 
the WLQs (e.g. radio-loudness, variability, and polarisation) are different from those of 
BL Lac objects
(McDowell et al. \cite{Mcdowell1995};
Shemmer et al. \cite{Shemmer2006};
Diamond-Stanic et al. \cite{Diamond2009};
Plotkin et al. \cite{Plotkin2010};
Lane et al. \cite{Lane2011};
Wu et al. \cite{Wu2012}).
WLQs are also different from type 2 quasars where only the broad line
components are missed in the unpolarised spectra, and the 
Eddington ratios $\varepsilon = L/L_{\rm Edd}$\footnote{The Eddington luminosity $L_{\rm Edd}$ is the
luminosity for the critical stable case where the gravitational pressure of the accretion flow is exactly
balanced by the pressure of the radiation flow.}
are usually lower 
(Tran et al. \cite{Tran2003}; 
Shi et al. \cite{Shi2010}; 
Shemmer et al. \cite{Shemmer2010}).
Factors that can mimic WLQ spectra are line absorption or a strong \ion{Fe}{ii} pseudo-continuum 
(e.g. Lawrence et al. \cite{Lawrence1988}). Such explanations may work for some objects but do
not explain the WLQ phenomenon in general (McDowell et al. \cite{Mcdowell1995}). Any scenario based 
on dust absorption has to be able to explain the weakening of the BELs 
without any reddening of the continuum.

Though a number of ideas have been developed, the WLQ phenomenon remains puzzling. 
The hypotheses can be roughly grouped into two families based on
either an extraordinary BLR or unusual properties of the central ionising source. 
The former includes, in particular, the ideas of 
generally abnormal properties of the BEL emitting clouds (Shemmer et al. \cite{Shemmer2010}),
a low covering factor of the BLR (i.e. a low fraction of the central source covered by 
BEL clouds, Niko{\l}ajuk et al. \cite{Nikolajuk2012}),
or a relative shortage of high-energy UV/X-ray photons due to a shielding gas with a high covering factor
that prevents the X-ray photons from reaching the BLR
(Lane et al. \cite{Lane2011};
Wu et al. \cite{Wu2012}).
Abnormal properties of the continuum source may include a high 
Eddington ratio  as in PHL 1811 
(e.g. Leighly et al. \cite{Leighly2007}; 
but see Hryniewicz et al. \cite{Hryniewicz2010};
Shemmer et al. \cite{Shemmer2010}),
a freshly launched wind from the accretion disk (Hryniewicz et al. \cite{Hryniewicz2010}),
an optically dull AGN (Comastri et al. \cite{Comastri2002}; Severgnini et al. \cite{Severgnini2003}), 
or a cold accretion disk around a high-mass ($M \ge  3\cdot 10^9 M_\odot$) black hole
(Laor \& Davis \cite{Laor2011}).

In a previous study (Meusinger et al. \cite{Meusinger2012}, hereafter Paper 1), we used 
the database of the $10^5$ quasar spectra from the SDSS Seventh Data Release (DR7, 
Abazajian et al. \cite{Abazajian2009}) to select the approximately one per cent  
of the quasars with the strongest deviations of their spectra from the ordinary quasar spectrum
as represented by the SDSS quasar composite spectrum (Vanden Berk et al. \cite{VandenBerk2001}).  
About one fifth of this sample was classified as uncommon because of remarkably 
weak BELs. We found that these WLQs are, on average, more luminous, have a steeper composite
spectrum (i.e. lower value of $\alpha_\lambda, F_\lambda \propto \lambda^{\alpha_\lambda}$) at 
$\lambda \ga 2000$\AA, and have a high percentage of radio-loud quasars (26\%). In addition, in a 
study of the variability of the quasars in the SDSS stripe 82 revealed, it was 
found that WLQs tend to have lower variability amplitudes (Meusinger et al. \cite{Meusinger2011}).  
No effort has been made to create complete WLQ samples in these previous studies.

The present paper is aimed at the construction and analysis of a more voluminous and more
thoroughly selected sample of WLQs and
WLQ-like objects by taking again advantage of the unprecedented spectroscopic data from the SDSS DR7. 
This study exploits the compilations of quasar properties by Schneider et al. (\cite{Schneider2010}) 
and Shen et al. (\cite{Shen2011}) that are based on the SDSS DR7 and include data from the 
1.4 GHz radio survey Faint Images of the Radio Sky at Twenty-Centimeters (FIRST; Becker et al. \cite{Becker1995}).
We apply essentially the same selection method as in Paper 1. The selection and the construction of the 
WLQ sample is described in Sect.\,\ref{sec:selection}.
Section\,\ref{sec:properties} is considered with the UV composite spectrum, the luminosities, 
black hole masses, Eddington ratios, accretion rates, and variability, where
particular attention is paid to the role of selection effects.  The wide band spectral energy 
distribution (SED) and the radio properties are the subject of Sect.\,\ref{sec:SED}.
The results are discussed in Sect.\,\ref{sec:discussion} and summarised in
Sect.\,\ref{sec:summary}.

%
%
\section{Selection of the WLQ sample}\label{sec:selection}
%
%

In Paper 1, we have shown that the selection of unusual spectra from the huge database of the 
SDSS can be performed efficiently by the combination of the power of the Kohonen self-organising map
(SOM) algorithm and the eyeball inspection of the resulting SOMs in the form of spectra icon maps. 
The Kohonen algorithm (Kohonen \cite{Kohonen2001}), an unsupervised learning process 
based on an artificial neural network, generates a low-dimensional 
(typically two-dimensional) map of complex input data. The SOM algorithm
and the software tool 
ASPECT\footnote{http://www.tls-tautenburg.de/fileadmin/research/meus/ASPECT/ ASPECT.html} 
used for the computation were described in detail in a separate paper 
(in der Au et al. \cite{inderAu2012}).

The present study is based on the 36 SOMs from Paper 1 for the nearly $10^5$ objects classified as quasars 
with $z = 0.6$ - 4.2 by the spectroscopic pipeline of the SDSS DR7. 
Each SOM consists of the quasars  within a $z$ interval of the width $\Delta z = 0.1$, sorted (clustered) 
according to the relative similarity of their spectra. For each SOM, an icon map was created 
where each object is represented by its SDSS spectrum with largely reduced spectral 
resolution. Despite the loss of resolution, the icon maps are well suited to quickly localise objects 
with special broad spectral features, such as unusually red or reddened continua, broad absorption lines (BALs), 
and unusually strong or unusually weak BELs. In Paper 1, we selected $10^3$ unusual 
spectra of different types. This selection was not primarily aimed at high completeness, and 
particularly the subsample of quasars with weak BELs was supposed to be substantially incomplete.
In the present study, we re-inspected all 36 spectra icon SOMs with the purpose to select solely
spectra with relatively weak BELs. Unlike Paper 1, the new selection includes only
spectra with clearly recognised quasar-typical spectral features that allows us to estimate the redshift; i.e.,
featureless blue spectra were not included here. In the first step, a total number of 
nearly $\sim2500$ candidates were selected as the initial sample of visually selected candidate WLQs,
among them are 2249 quasars listed in the SDSS DR7 quasar catalogue (Schneider et al. \cite{Schneider2010}).

The selection procedure is of course subjective and the selected sample is contaminated 
by other types of objects  such as quasars with line absorption (BALs or
associated narrow absorption lines) reducing the BEL flux, some early-type stars,
and also by different types of objects with noisy spectra where the signal-to-noise (S/N) ratio is 
too low for a certain classification.
To purify the initial sample, the selected objects were inspected individually 
in more detail by fitting the de-redshifted and  foreground extinction corrected SDSS spectra 
to the SDSS quasar composite spectrum from Vanden Berk et al. (\cite{VandenBerk2001}). 
The fitting algorithm is aimed to match both the positions of the typical quasar emission lines 
and the shape of the pseudo-continuum. The $z$ values derived in that way are generally in good agreement 
with those from the SDSS DR7 quasar catalogue, with only a few exceptions. Objects not identified in 
the quasar catalogue, as well as objects with discrepant $z$ values, were simply rejected from the sample.
The final WLQ sample consists of 365 quasars with counterparts in the SDSS DR7 quasar catalogue. 
The mean redshift is $\overline{z} = 1.50\pm0.45$. Figure\,\ref{fig:examples} displays six typical spectra.

\begin{figure}[bhtp]
\includegraphics[viewport=40 20 560 560,angle=270,width=8.0cm,clip]{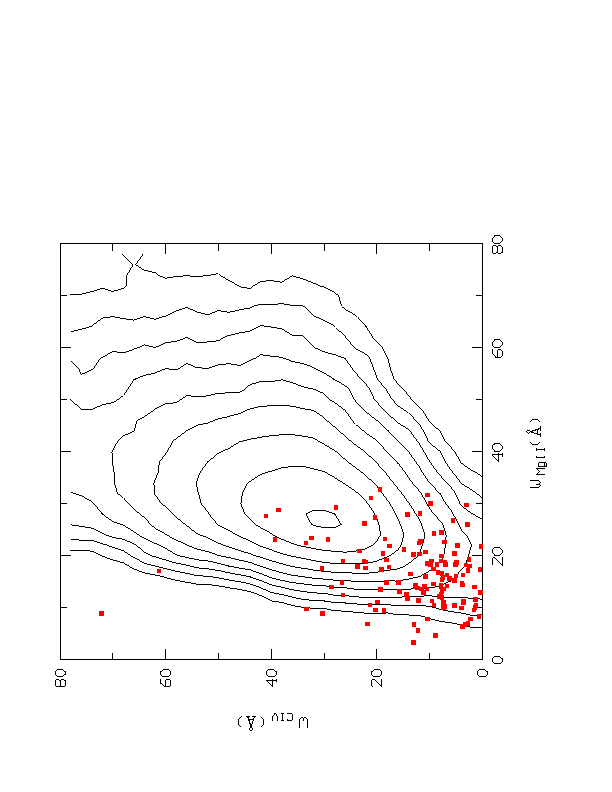}
\caption{Equivalent width $W_{\ion{C}{iv}}$ of the \ion{C}{iv} $\lambda$1549 line versus equivalent 
width $W_{\ion{Mg}{ii}}$ of the \ion{Mg}{ii} $\lambda$2798 line for the quasars from the
WLQ sample with $1.50\le z \le 2.22$ (filled red squares) and for 35137 quasars in the same redshift range 
from the Shen catalogue (equally spaced logarithmic local point density contours estimated with a 
grid size of $\Delta = 2$\AA\ on both axes). 
}
\label{fig:EWs}
\end{figure}

Relevant data for the WLQs were taken from Schneider et al. (\cite{Schneider2010})
and Shen et al. (\cite{Shen2011}). From the catalogue of quasar properties from SDSS DR7
(Shen et al. \cite{Shen2011}; hereafter Shen catalogue) we took in particular the equivalent widths (EWs) of the 
prominent BELs, the flux of the continuum close to these lines,
the bolometric luminosity $L$, the fiducial virial black hole mass $M$ from scaling relations, the Eddington 
ratio, and radio properties.

The line measurements in the Shen catalogue were made in the rest frame after removal of the 
Galactic foreground extinction. For each line, a local power law plus an iron template was fitted and
then subtracted from the spectrum. The resulting line spectrum was modelled by various functions.
Measurements of the EWs for both the \ion{Mg}{ii} $\lambda$2798 line and the \ion{C}{iv} $\lambda$1549 line 
are available for 136 of the selected WLQs with redshifts $1.50\le z \le 2.22$. 
Figure\,\ref{fig:EWs} shows the distribution of 
the WLQs on the $W_{\ion{Mg}{ii}}-W_{\ion{C}{iv}}$ plane in comparison with all
SDSS DR7 quasars in the same redshift interval. The centroid of the WLQs at 
$(W_{\ion{Mg}{ii}},W_{\ion{C}{iv}}) =$ (16.6$\pm$6.5\AA,14.2$\pm$14.6\AA) is clearly distant
from that of the ordinary quasars. On the other hand, our sample also contains quasars where 
obviously only one of the equivalent widths, $W_{\ion{Mg}{ii}}$ or $W_{\ion{C}{iv}}$, 
is low while the other is normal and even some quasars with normal values for both lines.
We did not reject these quasars to clean the WLQ sample further because a wider span of properties 
allows us to study some trends of other properties with the EW.

\begin{table}[hbpt]
\caption{32 objects rejected from the initial sample because of featureless spectra. The
type classification and the proper motion (pm) data were taken from the catalogues labelled in
the colums 3 and 6 and listed in the footnote to the table.
}
\begin{tabular}{lccrrc}
\hline
SDSS J             & Type   &  Ref. &  pm \ \ \ \ & $I_{\rm pm}$& Ref.\\
                   &        &       &(mas/yr)\\
\hline
003745.52+074423.2 & WD     &   1 &  44 &  5.1 & 9\\
021923.41+010413.4 & WD     &   1 &  18 &  2.2 & 9\\
024058.79-003934.5 & WD     &   1 &  23 &  2.7 & 9\\
073249.49+354651.5 & WD     &   1 &  40 &  5.6 & 9\\
084732.74+172819.1 & WD     &   2 &  70 &  9.9 & 9\\
084749.21+183016.8 & WD     &   1 & 151 & 21.4 & 9\\
085246.87+100523.0 & WD     &   1 &  61 &  5.9 & 9\\
093958.63+340152.4 & pm star&   3 &  58 &  8.4 & 9\\
094857.88+123243.0 & WD     &   1 & 188 & 23.4 & 9\\
095933.22+144548.9 & WD     &   4 &  70 &  9.3 & 9\\
100149.22+144123.8 & WD     &   5 & 346 & 90.7 & 5\\
101509.57+351813.8 & WD     &   1 &  65 &  8.3 & 9\\
105430.62+221054.8 & BL Lac &  6,7&   4 &  0.7 & 9\\
113245.62+003427.7 & BL Lac &  6,7&   2 &  0.4 & 9\\
120423.80+230913.3 & WD     &   1 &  51 &  7.8 & 9\\
121856.69+414800.2 & WD     &   1 &  34 &  4.6 & 9\\
122008.29+343121.7 & BL Lac &  6,7&   8 &  1.3 & 9\\
124510.00+570954.3 & BL Lac &  6,7&   5 &  0.9 & 9\\
124818.78+582028.9 & BL Lac &  6,7&   6 &  0.9 & 9\\
125335.04+163020.5 & WD     &   1 & 139 & 18.5 & 9\\
130210.74+454424.3 & pm star&   3 &  46 &  6.2 & 9\\
132232.05+373032.9 & WD     &   1 &  49 &  7.4 & 9\\
133040.69+565520.1 & BL Lac &  6,7&   6 &  0.6 & 9\\
141904.67+110306.2 & WD     &   4 &  19 &  2.6 & 9\\
145427.13+512433.7 & BL Lac &  6,7&   5 &  0.8 & 9\\
152913.56+381217.5 & BL Lac &  6,7&   4 &  0.7 & 9\\
153324.26+341640.3 & BL Lac &  6,7&   4 &  0.8 & 9\\
160410.22+432614.6 & WLQ    &   7 &   9 &  1.4 & 9\\
161315.36+511608.3 & WD     &   1 & 112 &  6.6 & 9\\
170108.89+395443.0 & BL Lac &  6,7&   5 &  0.7 & 9\\
220911.31-003543.0 & WD     &   1 &   6 &  0.7 & 9\\
224303.81+221456.0 & CV     &   8 &   7 &  1.0 & 9\\
\hline
\end{tabular}

\vspace{3mm}
{\bf References.}
1 - Kleinman et al. (\cite{Kleinman2013});
2 - McCook \& Sion (\cite{McCook1999});
3 - SDSS DR10 explorer;
4 - Girven et al. (\cite{Girven2011});
5 - Paper 1 (Tab.\,2);
6 - Massaro et al. (\cite{Massaro2009}, catalogue version August 2012);
7 - Plotkin et al. (\cite{Plotkin2010});
8 - Thostensen \& Skinner (\cite{Thorstensen2012});
9 - R\"oser et al. (\cite{Roeser2010})
\label{tab:featureless}
\end{table}

As mentioned above, the WLQ selection was confined to spectra with clearly recognisable quasar-typical
features. The advantage of reliable redshifts is however at the expense of the completeness of such a sample
at low EWs. In particular, we rejected 32 objects with blue spectra from our initial sample
because the spectra appeared more or less featureless to us. To estimate how strongly our final sample may
be biased against low-EW WLQs, we checked, a posteriori, several catalogues for additional information
on these 32 rejected objects (Tab.\,\ref{tab:featureless}).
Altogether 18 objects were classified as white dwarfs (WDs) of spectral
type DC or DQ, another one as a cataclysmic variable (CV). Ten sources are classified as high-confidence
radio-loud BL Lacertae objects in the  catalogue of optically selected BL Lac objects
from the SDSS DR7 (Plotkin et al. (\cite{Plotkin2010}) and also in the Roma-BZCAT
multifrequency catalogue of blazers (Massaro et al. \cite{Massaro2009}).
One object, SDSS J160410.22+432614.6, is listed in the Plotkin et al. catalogue of weak-featured
radio-quiet objects and may be a WLQ that is missed in the present sample. For the remaining two objects, 
the SDSS DR10 explorer tool gives high proper motions suggesting that they are nearby stars, most likely WDs.

Absolute proper motions are very helpful in distinguishing extragalactic
objects from nearby white dwarfs. Table\,\ref{tab:featureless} lists the proper motions
taken from the PPMXL catalogue (R\"oser et al. \cite{Roeser2010}). In addition, a proper motion index,
$I_{\rm pm}$ is given that is defined as the pm in units of the pm error.
For the 10 BL Lac objects, the mean pm index $I_{\rm pm} = 0.8$ indicates zero proper motion.
In contrast, the sample of the 18 WDs plus one CV has a mean pm of 79 mas/yr and a mean pm index
$I_{\rm pm} = 12.9$. The inspection of the individual sources yields that the pm index does not indicate
significant proper motions for all BL Lac objects, the probably missed WLQ SDSS J160410.22+432614.6, and for
two objects classified as stellar in the literature (SDSS J220911.31-003543.0 and SDSS J224303.81+221456.0).
A low pm does not necessarily exclude that these objects are nearby stars. On the other hand, however,
we cannot definitively exclude that they are WLQs.

Only three objects from Tab.\ref{tab:featureless} are in the Shen catalogue: the probable WLQ
SDSS J160410.22+432614.6 and the two BL Lac objects SDSS J113245.62+003427.7 and
SDSS J170108.89+395443.0. The EWs of the \ion{Mg}{ii} line are
1.0\,\AA, 5.8\,\AA, and 9.4\,\AA. The mean value of 5.4\,\AA\ is close to the 5\,\AA\ EW limit 
for the selection of BL Lac objects (e.g. Plotkin et al. \cite{Plotkin2008}; Ghisellini
et al. \cite{Ghisellini2011}). The EWs of the WLQ sample follow a log-normal distribution with the lower 
3$\sigma$ deviation from the mean value at 4.6\,\AA. There are only two WLQs (0.5\%) with lower EWs
in our sample. We assume therefore that the minimum \ion{Mg}{ii} EW threshold of our selection is at 
about 5\,\AA. Given that this threshold is exceeded by SDSS J170108.89+395443.0 with $W_{\ion{Mg}{ii}} = 9.4$\,\AA,
we conclude that Tab.\,\ref{tab:featureless} may contain up to altogether four wrongly rejected WLQs.

For some purposes it is useful to restrict the quasar sample to the redshift interval $0.7 \le z \le 1.7$
(Meusinger \& Weiss \cite{Meusinger2013}). First, the formal uncertainties of the 
catalogued bolometric luminosities and black hole masses are lowest for such redshifts.
As pointed out by Shen et al. (\cite{Shen2011}), the mass uncertainty was
propagated from the measurement uncertainties of the continuum luminosities and the line width, but does
neither include systematic effects nor the statistical uncertainty from the calibration of the 
scaling relations. Secondly, in that $z$ interval, the fiducial black hole masses from the Shen catalogue
are uniformly derived from the scaling relation for the \ion{Mg}{ii} line. 
Virial masses derived from the \ion{C}{iv} line are considered to be less reliable than those from
the H$\beta$ or the \ion{Mg}{ii} line (e.g. Shen \& Liu \cite{Shen2012}).
Moreover, the estimation 
of the accretion rate (Sect.\,\ref{subsec:lum_etc}) requires the knowledge of the continuum 
slope $\alpha_\lambda$ between 3000 and 4800\AA, which can be measured from the SDSS spectra 
only for not too high $z$. The restricted WLQ sample (hereafter: rWLQ sample)  with $0.7 \le z \le 1.7$ contains 
261 quasars. It is particularly useful when luminosities, black hole masses, Eddington ratios,
or accretion rates are considered.

Because it is difficult to compare the resulting WLQ sample with known WLQ samples defined
by the measured EWs, it is helpful to consider a subsample that follows a 
statistically based selection. Following Diamond-Stanic et al. (\cite{Diamond2009}), we assumed a log-normal EW 
distribution and defined WLQs as quasars having EWs below a 3$\sigma$ threshold. 
For the quasars from the Shen catalogue in  the redshift range of the rWLQ sample 
we obtain the mean EWs 34.1\,\AA,  38.7\,\AA\
and the WLQ selection thresholds 11\,\AA, 4.8\,\AA\ for the \ion{Mg}{ii} line and the \ion{C}{iv} line,
respectively. 
The subsamples from the entire WLQ sample and the rWLQ sample, respectively, with EWs below these 
thresholds are hereafter indicated by the suffix EWS. 
The WLQ-EWS subsample consists of 46 quasars with a mean redshift $\overline{z} = 1.48\pm0.56$ and mean 
EWs $8.8\pm1.9$\,\AA, $2.6\pm1.2$\,\AA.
Assuming that Tab.\,\ref{tab:featureless} contains at most four wrongly rejected WLQs, the completeness of
this sample, compared with other samples based on the SDSS DR7 spectroscopic quasar catalogue, is expected to be
about 90\%. It must be noted, however, that this catalogue itself is biased against quasars with very
week emission lines. The rWLQ-EWS subsample consists of 33 quasars with $\overline{z} = 1.24\pm0.24$ and
$9.0\pm1.8$\,\AA, $2.5\pm1.3$\,\AA.

\vspace{0.2cm}
\begin{figure}[htbp]
\begin{centering}
\includegraphics[viewport=95 0 555 790,angle=270,width=9.0cm,clip]{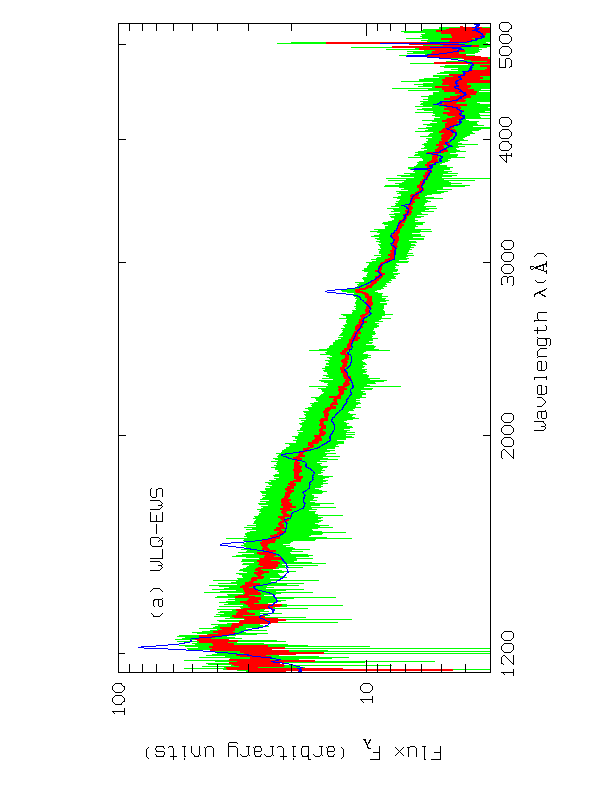}
\includegraphics[viewport=85 0 555 790,angle=270,width=9.0cm,clip]{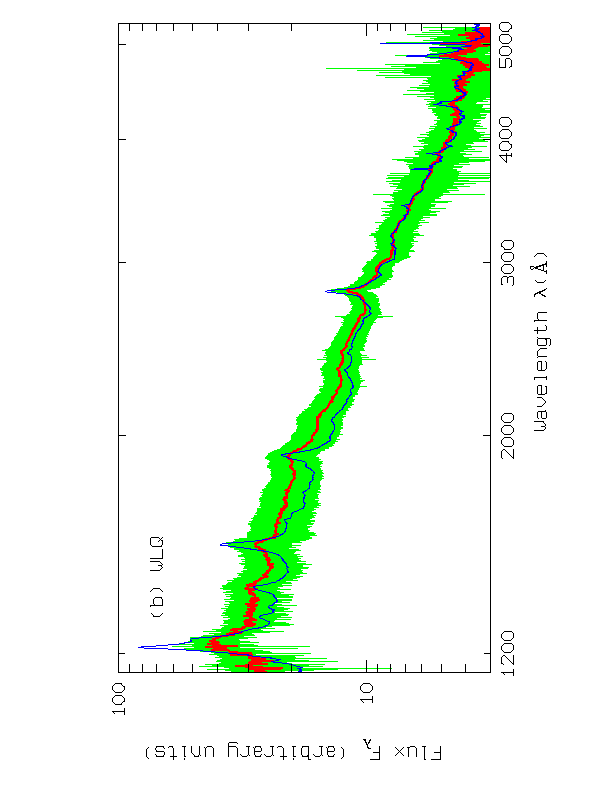}
\includegraphics[viewport=85 0 580 790,angle=270,width=9.0cm,clip]{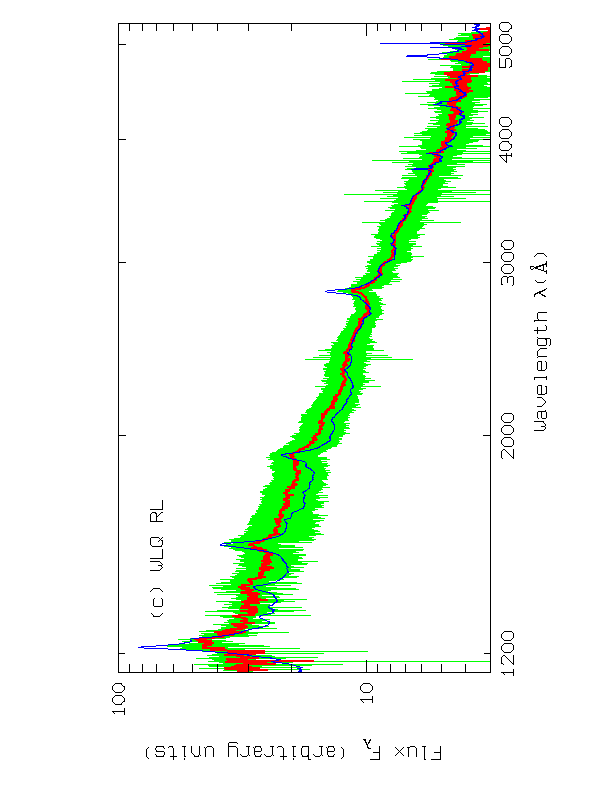}
\end{centering}
\caption{Arithmetic median composite spectrum (thick red curve) and standard deviation (green shaded area) 
for (a) the WLQ-EWS sample, (b) the entire WLQ sample, and (c) the radio-loud subsample. 
For comparison, the SDSS quasar composite
spectrum (thin blue curve) from Vanden Berk et al. (\cite{VandenBerk2001}) is over-plotted,
normalised to match the WLQ composites at the right hand side.}
\label{fig:composite}
\end{figure}

%
%
\section{UV/optical properties of the selected quasars}\label{sec:properties}
%
%

In this Section, we analyse mean properties of our WLQs where various subsamples are considered.
Particular attention should be payed to the WLQ-EWS and rWLQ-EWS subsamples because they are closest to traditional 
WLQ samples. On the other hand, these subsamples are small and it thus makes sense to also consider the 
corresponding parent samples for comparison. In the context of the black hole mass, the Eddington ratio, and 
especially the accretion rate, the rWLQ-EWS and the rWLQ sample should be preferred (see above). 
For the comparison with normal quasars, the $z$ dependence of quasar properties
must be taken into consideration because the $z$ distribution of the rWLQ sample is different from that of 
the entire Shen catalogue.  Therefore, we created a comparison sample of ordinary quasars with the
same $z$ distribution following the procedure outlined in Paper 1. The comparison sample is about ten times
larger than the rWLQ sample. For the mean EW of the comparison quasars we find 
$(W_{\ion{Mg}{ii}}, W_{\ion{C}{iv}}) =$ (42$\pm$22 \AA, 51$\pm$52 \AA) compared with 
(17$\pm$7 \AA, 17$\pm$18 \AA) for the rWLQ sample (Tab.\,\ref{tab:mean-prop}).

\subsection{Composite spectra}\label{subsec:composite} 

The individual inspection of the SDSS spectra led to the impression that many WLQs show a steeper continuum 
than the SDSS quasar composite spectrum (see Fig.\,\ref{fig:examples}). 
This tendency was already reported in Paper 1, but it appears even more pronounced in the present WLQ sample.

We computed composite spectra in the same way as in Paper 1. The procedure is essentially based
on the combining technique described by Vanden Berk et al. (\cite{VandenBerk2001}).  
Fig.\,\ref{fig:composite} shows the resulting arithmetic median composite spectra for 
(a) the WLQ-EWS subsample and (b) the entire WLQ sample.
For comparison, the SDSS quasar composite spectrum from 
Vanden Berk et al. (\cite{VandenBerk2001}) is over-plotted, arbitrarily normalised at 3200 \AA. The comparison 
clearly illustrates the weaker BELs and the steeper  (i.e. less reddened) ultraviolet continuum
for both (a) the WLQ-EWS subsample and (b) the entire WLQ sample.

We also confirm the finding from Paper 1 that the WLQ spectra flatten at short wavelengths, 
approximately below 2000 \AA. We estimated the spectral slopes of the individual spectra by 
fitting the $F_\lambda \propto \lambda^{\alpha_\lambda}$ 
power law in pseudo-continuum windows (see Paper 1) and found a mean value of
$\overline{\alpha_\lambda}=-1.69\pm0.36$ for the rWLQs compared to $-1.53\pm0.39$ for the ordinary 
quasars from the comparison sample with S/N$>5$ in both cases.
For the  97 quasars with $1.7 < z < 3$ in the full WLQ sample the mean slope is 
$\overline{\alpha_\lambda}=-1.52\pm0.35$.

A possible explanation of the steeper continuum could be an additional non-thermal component in WLQ spectra. 
Therefore, we also computed the composite spectrum of the subsample of radio-loud WLQ. Usually, quasars 
are classified as  radio-loud based on the 
radio-to-optical flux ratio. Kellermann et al. (\cite{Kellermann1989}) defined the radio-loudness parameter 
$R = F_{\rm 5GHz}/F_{\rm 4400\AA}$, where $F_{\rm 5GHz}$ and $F_{\rm 4400\AA}$ are the 
flux densities at rest-frame 5 GHz and 4400\AA, 
and called quasars with $R$ greater than 10 as radio-loud. Since then,
$R=10$ is commonly used to distinguish radio-loud and radio-quiet quasars
(e.g. 
Francis et al. \cite{Francis1993};
Urry \& Padovani \cite{Urry1995};
Ivezi\'c et al. \cite{Ivezic2002};
McLure \& Jarvis \cite{McLure2004};
Richards et al. \cite{Richards2011}),
although this value is to some degree rather arbitrary 
(e.g. Falcke et al. \cite{Falcke1996}; 
Wang et al. \cite{Wang2006};
Sect.\,\ref{sec:discussion} below).
We took the radio-loudness parameter from the Shen catalogue. Shen et al. (\cite{Shen2011}) estimated 
the radio-loudness $R = F_{\rm 6cm}/F_{\rm 2500\AA}$ following Jiang et al. (\cite{Jiang2007}), where
$F_{\rm 6cm}$ is the flux density at rest-frame 6\,cm determined from 
the FIRST integrated flux density at 20\,cm assuming a power law $F_\nu \propto \lambda^{\alpha_\nu}$  with 
$\alpha_\nu = -0.5$. The rest-frame
2500\,\AA\ flux density $F_{\rm 2500\AA}$ is determined from the power-law continuum fit to the SDSS spectrum.
Following Jiang et al. (\cite{Jiang2007}), we use the criterion $R>10$ to classify radio-loud quasars.
Figure\,\ref{fig:composite}\,c shows the radio-loud WLQ composite spectrum.
It is very similar to that of the entire WLQ sample, which is dominated by radio-quiet quasars. We conclude 
that there is no significant difference between the composites of radio-loud and of radio-quiet WLQs, in 
accordance with earlier findings for ordinary quasars (e.g. Francis et al. \cite{Francis1993}).

\vspace{0.2cm}
\begin{figure}[htbp]
\begin{centering}
\includegraphics[viewport=30 20 560 795,angle=270,width=8.8cm,clip]{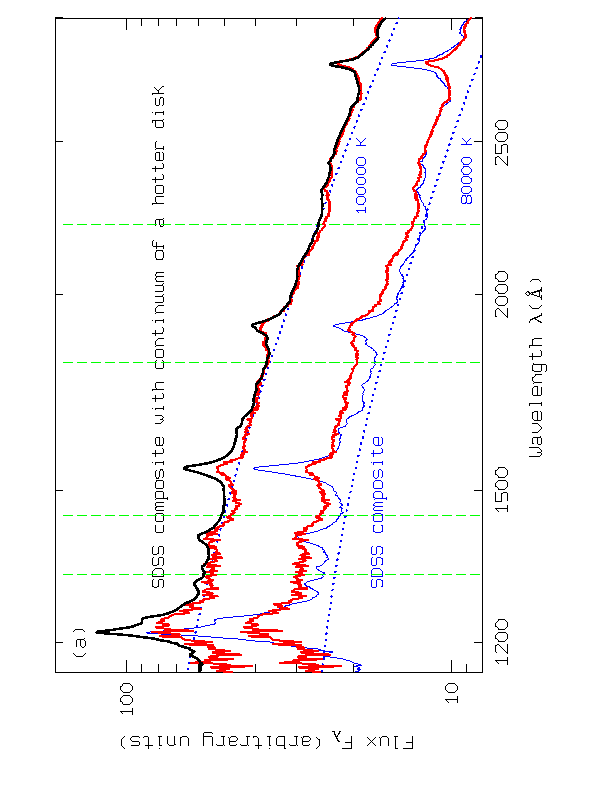}
\includegraphics[viewport=30 20 570 795,angle=270,width=8.8cm,clip]{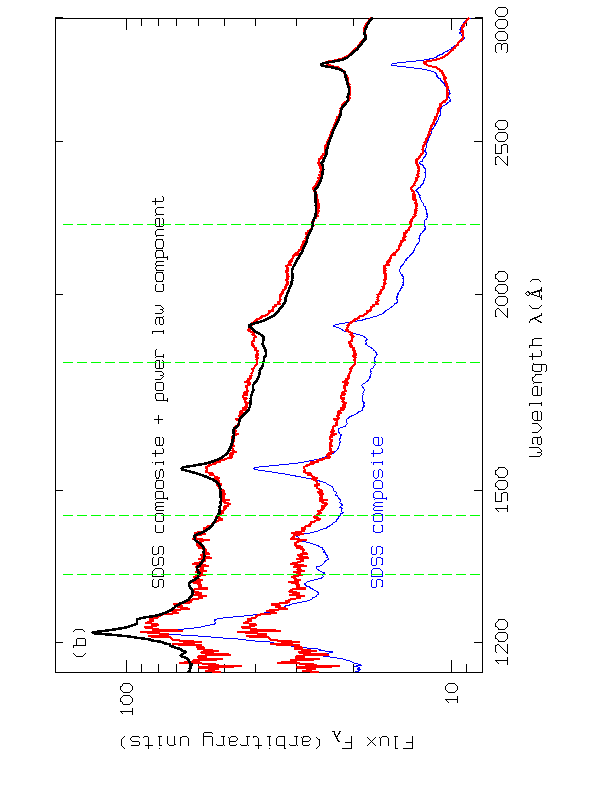}
\end{centering}
\caption{
Two interpretations of the steeper slope of the WLQ continuum.
(a): At the bottom of the panel, the WLQ composite spectrum (red) is compared with the slightly
smoothed SDSS quasar composite spectrum (solid blue) fitted by a MTBB model with temperature parameter
$T^\ast =  8\,10^4$\,K (dotted blue).
The thick black curve above is a modified version of the SDSS quasar composite with a hotter continuum
fitted to the MTBB model for $T^\ast = 10^5$\,K and compared with the WLQ composite spectrum (red) that
is normalised to match the former at the right hand side.
(b): As (a), but the modification of the SDSS quasar composite spectrum consists of an additional
power-law component instead of a hotter continuum. Vertical lines: continuum windows.
}
\label{fig:fit}
\end{figure}

It is known from both multi-epoch photometry and spectroscopy of SDSS quasars that the variability in the 
emission line flux is only $\sim10\%$ of the variability in the underlying continuum  
(Wilhite et al. \cite{Wilhite2005}; Meusinger et al. \cite{Meusinger2011}). 
Let us assume that the WLQs represent a stage where the continuum is significantly enhanced but 
the lines are not. 
At the bottom of Fig.\,\ref{fig:fit}\,a, the continuum of the SDSS composite spectrum
between the Ly$\alpha$ and the \ion{Mg}{ii} line is fitted by a simple multi-temperature black body (MTBB) model
with a temperature parameter $T^\ast \approx 8\, 10^4$\,K that provides a good fit to the observed quasar
composite spectra over a much wider wavelength range (e.g. Meusinger \& Weiss 2013).
The steeper WLQ composite requires a higher temperature. 
Indeed, the fit is considerably improved when we subtract the continuum of the $8\cdot 10^4$\,K MTBB from 
the SDSS composite and add instead the continuum of the $10^5$\,K model.
An alternative way to achieve a similar result is shown in Fig.\,\ref{fig:fit}\,b.
We simply added a hypothetical further power-law component $F_\lambda \propto \lambda^{-1.7}$, 
i.e. $\alpha_\nu = -0.3$, of approximately the same level as the thermal continuum at 3000\,\AA. 
In both cases, the WLQ composite spectrum is well matched by the modified SDSS quasar composite spectrum. 
The enhancement of the continuum flux dilutes the line flux and reduces the EWs of the lines
correspondingly.

\subsection{Luminosity, black hole mass, Eddington ratio, and accretion rate}\label{subsec:lum_etc}

\subsubsection{Mean properties and correlation diagrams}

Table\,\ref{tab:mean-prop} lists the mean values of the bolometric luminosity $L$, the black hole mass $M$,
the Eddington ratio $\varepsilon$, and the accretion rate $\dot{M}$ 
for the WLQ-EWS subsample, the rWLQ-EWS subsample, the corresponding parent samples,
and the comparison sample. The rWLQ sample is subdivided into the subsamples of radio-loud 
and not radio-loud quasars, and the radio-loud subsample is subdivided into core-dominated and lobe-dominated 
radio sources. Quasars with $R>10$ are classified as radio-loud (Sect.\,\ref{subsec:composite}). 
The not radio-loud subsample 
includes both radio-quiet quasars, i.e. $R<10$, and quasars outside the FIRST footprint area. 
With the exception of $\dot{M}$, all data were taken from the Shen catalogue. 
The distinction of lobe-dominated and core-dominated radio sources is based on the parameter
{\sc first\_radio\_flag}.

\begin{table*}[hbpt]
\caption{
Mean properties of the WLQ, rWLQ, WLQ-EWS, rWLQ-EWS, and comparison sample.
}
\centering
\begin{tabular}{lrcccccccc}
\hline
Sample  &
Number  &
Mean $z$ &
$W_{\rm \ion{Mg}{ii}}$ (\AA)&
$W_{\rm \ion{C}{iv}}$ (\AA) &
$L (10^{46}$erg/s)          &
$M (10^9 M_\odot)$          &
$\varepsilon$               &
$\dot{M} (M_\odot/$yr) \\
\hline
WLQ-EWS\tablefootmark{a}  &  46&$1.48\pm0.56$&$  9\pm2$&$ 3\pm1  $&$5.5\pm3.6$ &$1.8\pm1.8$&$0.47\pm0.46$& - \\
rWLQ-EWS\tablefootmark{a} &  33&$1.24\pm0.24$&$  9\pm2$&$ 3\pm1  $&$4.9\pm3.5$ &$1.7\pm1.8$&$0.38\pm0.29$&$7.6\pm6.4$\\
WLQ all                   & 365&$1.50\pm0.45$&$ 17\pm8$&$13\pm14 $&$7.7\pm11.2$&$3.1\pm4.1$&$0.32\pm0.30$& - \\
rWLQ all                  & 261&$1.32\pm0.24$&$ 17\pm7$&$17\pm18 $&$5.7\pm5.7$ &$2.7\pm2.9$&$0.28\pm0.22$&$6.5\pm7.0$\\
rWLQ nRL\tablefootmark{b} & 202&$1.32\pm0.24$&$ 18\pm7$&$17\pm17 $&$5.3\pm5.2$ &$2.5\pm2.9$&$0.27\pm0.21$&$6.0\pm6.9$\\
rWLQ RL\tablefootmark{c}  &  59&$1.31\pm0.22$&$ 14\pm6$&$20\pm23 $&$7.3\pm7.0$ &$3.1\pm2.9$&$0.30\pm0.26$&$8.0\pm7.0$\\
rWLQ RL,c\tablefootmark{d}&  49&$1.33\pm0.22$&$ 14\pm6$&$20\pm24 $&$7.2\pm7.3$ &$2.9\pm2.8$&$0.32\pm0.27$&$7.9\pm7.1$\\
rWLQ RL,l\tablefootmark{e}&  10&$1.22\pm0.21$&$ 17\pm5$&$22\pm00 $&$7.7\pm5.5$ &$4.3\pm2.9$&$0.22\pm0.18$&$8.3\pm7.0$\\
Comparison                &2750&$1.32\pm0.24$&$42\pm22$&$51\pm52 $&$1.5\pm1.6$ &$1.1\pm1.1$&$0.17\pm0.17$&$2.6\pm3.3$\\
\hline
\end{tabular}
\tablefoot{
\tablefoottext{a}{equivalent width selected subsample with $W_\ion{Mg}{ii}<11$\AA,$W_\ion{C}{iv}<4.8$\AA;}
\tablefoottext{b}{not radio-loud;}
\tablefoottext{c}{radio-loud;}
\tablefoottext{d}{radio-loud and core-dominated;}
\tablefoottext{e}{radio-loud and lobe-dominated.}
}
\label{tab:mean-prop}
\end{table*}

\begin{table*}[hbpt]
\caption{Test statistics $D_{\rm max}$ from the Kolmogorov-Smirnov two sample test for the comparison of the
first six samples from Tabl.\,\ref{tab:mean-prop} with the comparison sample.
{\bf In the last two columns},
the critical values $D_{\rm crit,\alpha}$ of the one-tailed test are given for
the error probabilities $\alpha =$ 0.01 and $\alpha =$ 0.001.}
\centering
\begin{tabular}{lcccccc}
\hline
Sample   &
$D\,(\log L)$      &
$D\,(\log M)$     &
$D\,(\varepsilon)$  &
$D\,(\dot{M})$   &
$D_{\rm crit,0.01}$ &
$D_{\rm crit,0.001}$ \\
\hline
WLQ-EWS  & 0.67  & 0.20  & 0.52  & -     & 0.23 & 0.28\\
rWLQ-EWS & 0.62  & 0.21  & 0.50  & 0.55  & 0.27  & 0.33\\
WLQ all  & 0.66  & 0.32  & 0.33  & -     & 0.09  & 0.10\\
rWLQ all & 0.62  & 0.31  & 0.30  & 0.40  & 0.10  & 0.12\\
rWLQ nRL & 0.59  & 0.28  & 0.28  & 0.39  & 0.11  & 0.14\\
rWLQ RL  & 0.69  & 0.42  & 0.38  & 0.49  & 0.20  & 0.24\\
rWLQ RL,c& 0.70  & 0.37  & 0.39  & 0.50  & 0.22  & 0.27\\
\hline
\end{tabular}
\label{tab:KS-test}
\end{table*}

We estimated the accretion rate in the same way as in Meusinger \& Weiss (\cite{Meusinger2013}). 
This approach is based on the scaling relation for $\dot{M}$ derived by Davis \& Laor (\cite{Davis2011}) 
from the standard accretion disk model. As was shown by Davis \& Laor (\cite{Davis2011}),
this relation can be used to compute $\dot{M}$ from the optical luminosity and is thus less affected 
by the uncertainties of the innermost part of the accretion disk. We slightly modified this relation to
\begin{equation}\label{eqn:AR_DL}
 \dot{M} = 3.5\cdot 0.64^{-1.5\cdot(1+\alpha_\lambda)}\, L_{3000, 45}^{1.5}\, M_8^{-0.89},
\end{equation}
by extrapolating the optical luminosity from the continuum luminosity $L_{3000}$ at 3000 \AA\,
adopting a power law with the spectral index $\alpha_\lambda$ where $L_{3000, 45}$  is
$L_{3000}$ in units of $10^{45}\,\mbox{erg}\,\mbox{s}^{-1}$
and $M_8 $ is the black hole mass in units of $10^8 M_\odot$.
We derived the continuum spectral index $\alpha_\lambda$  for each quasar individually
by fitting a power law to the foreground extinction-corrected SDSS spectrum in
the continuum windows. These individual slopes were used for all those
quasars having spectra with $S/N>3$ in at least three windows. For quasars with noisier spectra we simply
adopted the mean slope $\alpha_\lambda = -1.52$ from the composite spectrum of the parent sample.
The average accretion rate for the rWLQ-EWS sample is three times higher than for the
comparison sample. The rWLQ sample shows a higher average accretion rate as well.\footnote{The accretion 
rates estimated for the WLQ-EWS sample and the entire WLQ sample, respectively, are even higher, but are 
not listed in Tab.\,\ref{tab:mean-prop} because of the anticipated higher uncertainties.}

\begin{figure*}[htbp]
\begin{tabbing}
\includegraphics[viewport=0 52 550 560,angle=0,width=6.0cm,clip]{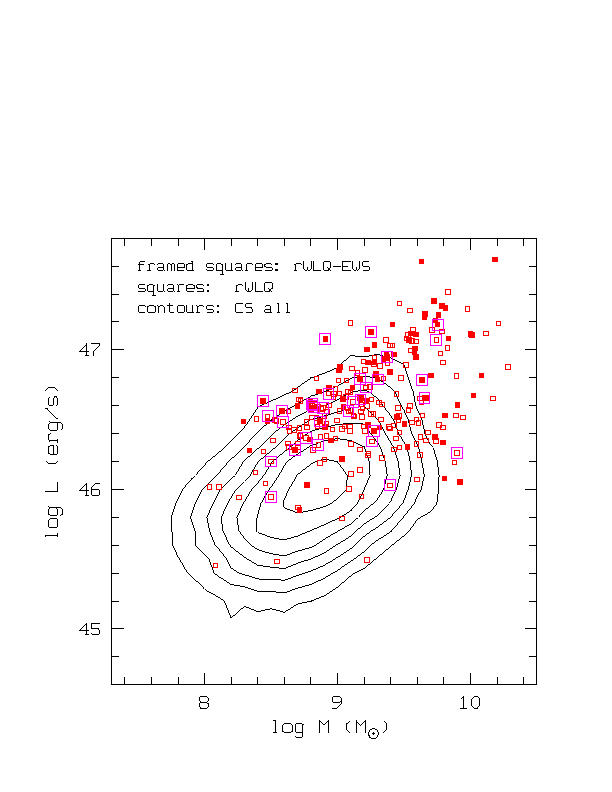}\hfill
\includegraphics[viewport=0 52 550 560,angle=0,width=6.0cm,clip]{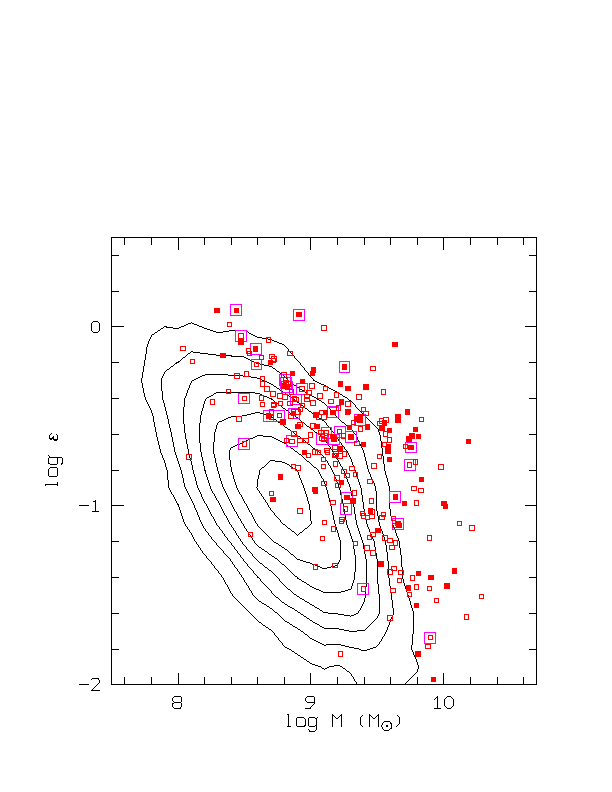}\hfill
\includegraphics[viewport=0 52 550 560,angle=0,width=6.0cm,clip]{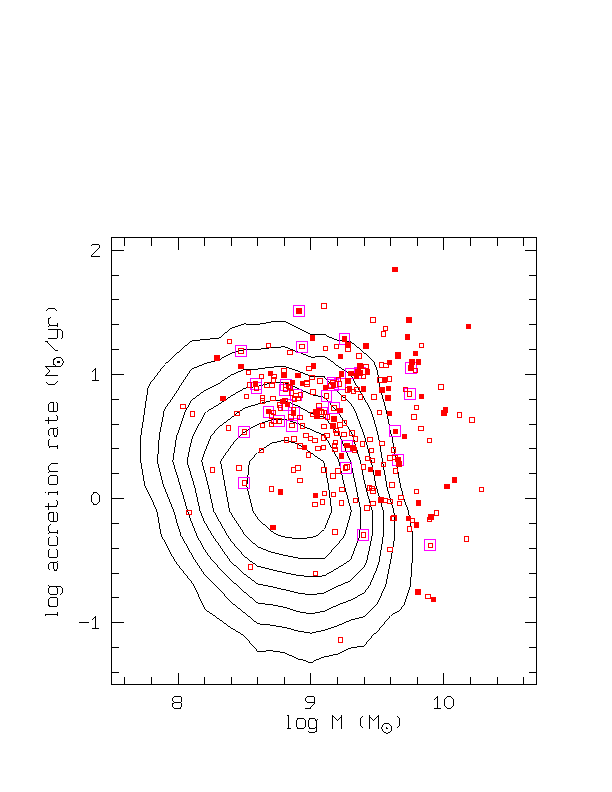}\hfill \\
\includegraphics[viewport=0 52 550 560,angle=0,width=6.0cm,clip]{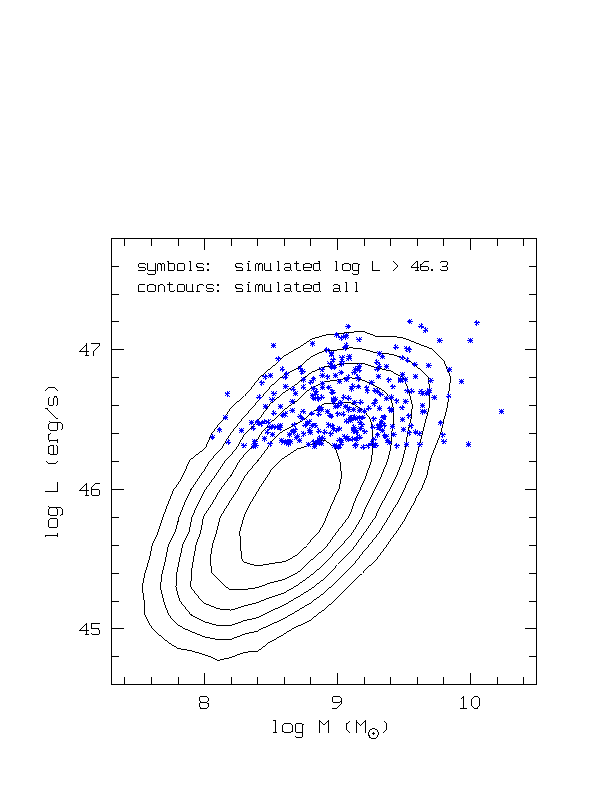}\hfill
\includegraphics[viewport=0 52 550 560,angle=0,width=6.0cm,clip]{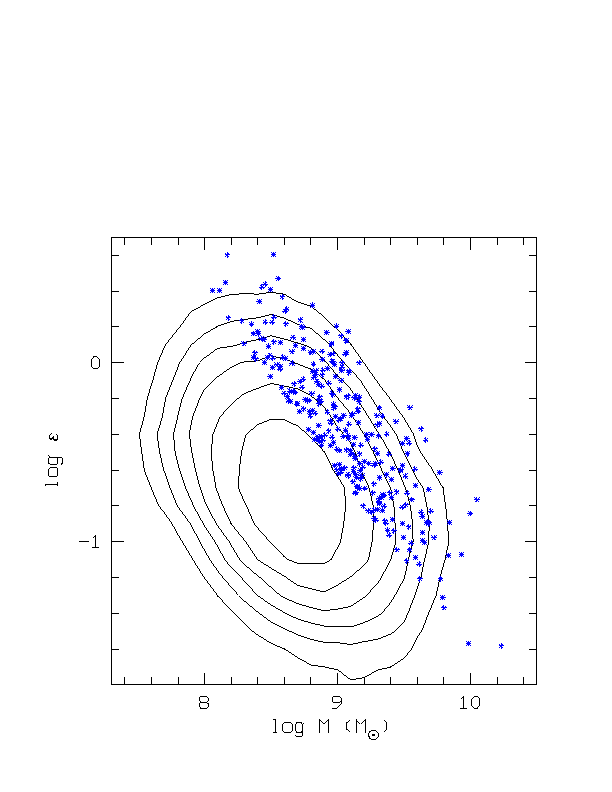}\hfill
\includegraphics[viewport=0 52 550 560,angle=0,width=6.0cm,clip]{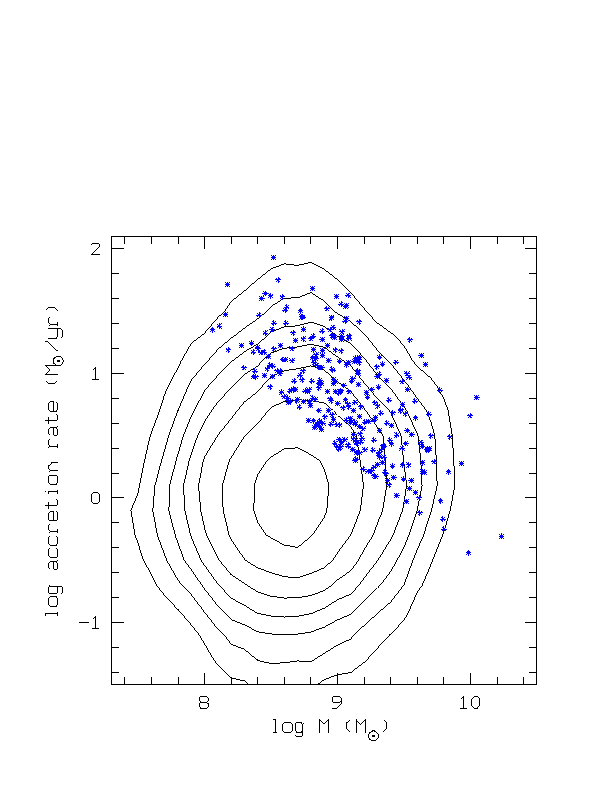}\hfill \\
\includegraphics[viewport=0 52 550 560,angle=0,width=6.0cm,clip]{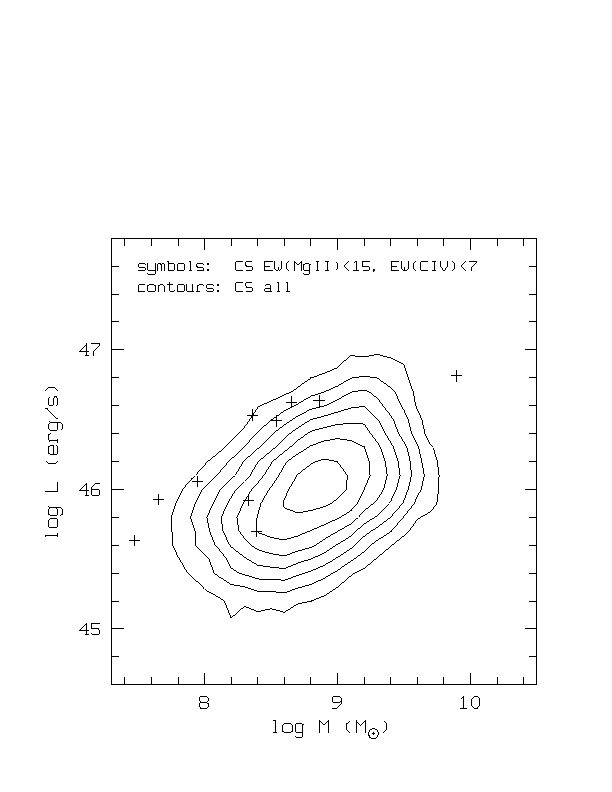}\hfill
\includegraphics[viewport=0 52 550 560,angle=0,width=6.0cm,clip]{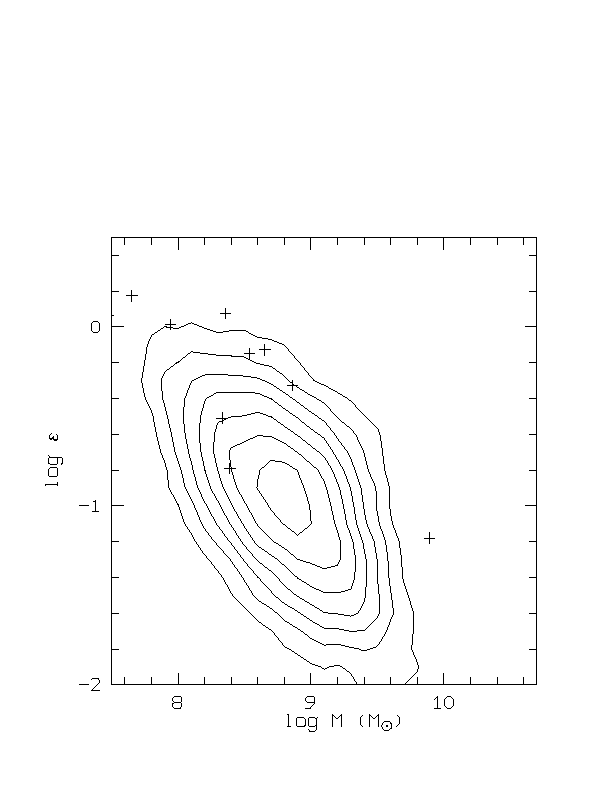}\hfill
\includegraphics[viewport=0 52 550 560,angle=0,width=6.0cm,clip]{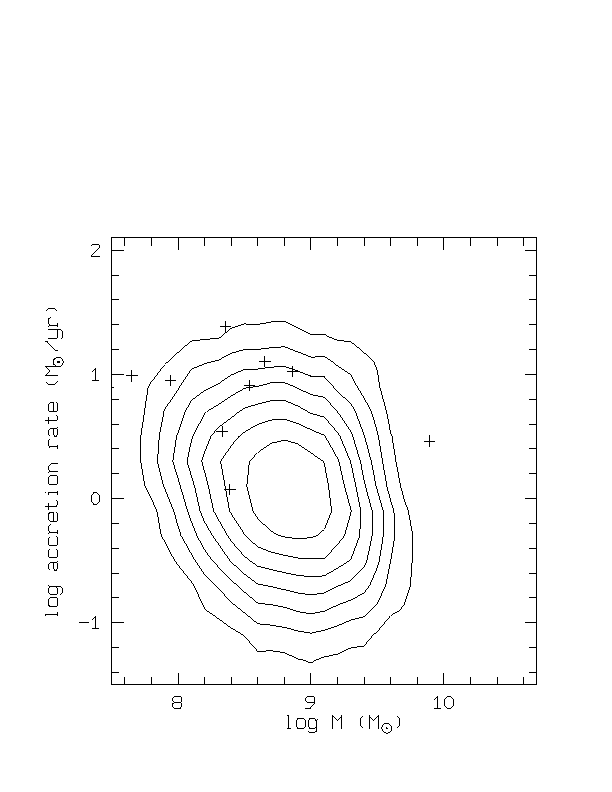}\hfill \\
\includegraphics[viewport=0 52 550 560,angle=0,width=6.0cm,clip]{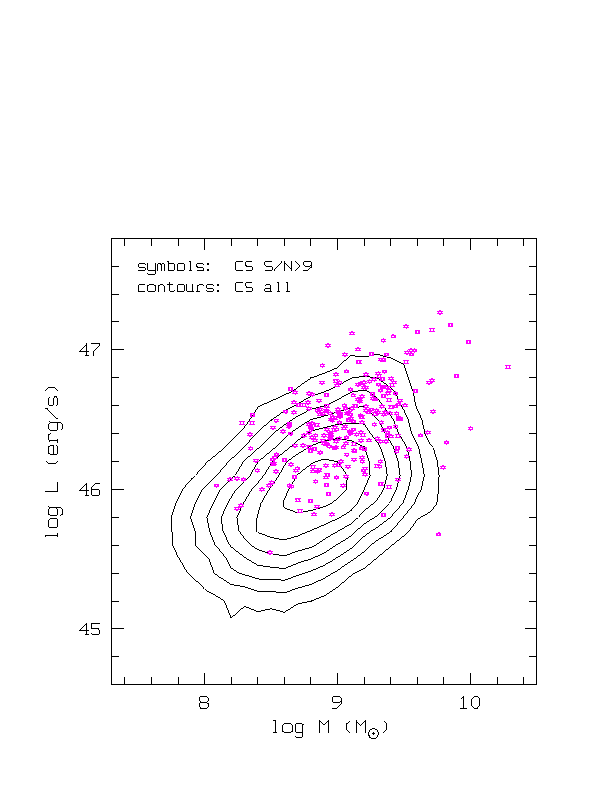}\hfill
\includegraphics[viewport=0 52 550 560,angle=0,width=6.0cm,clip]{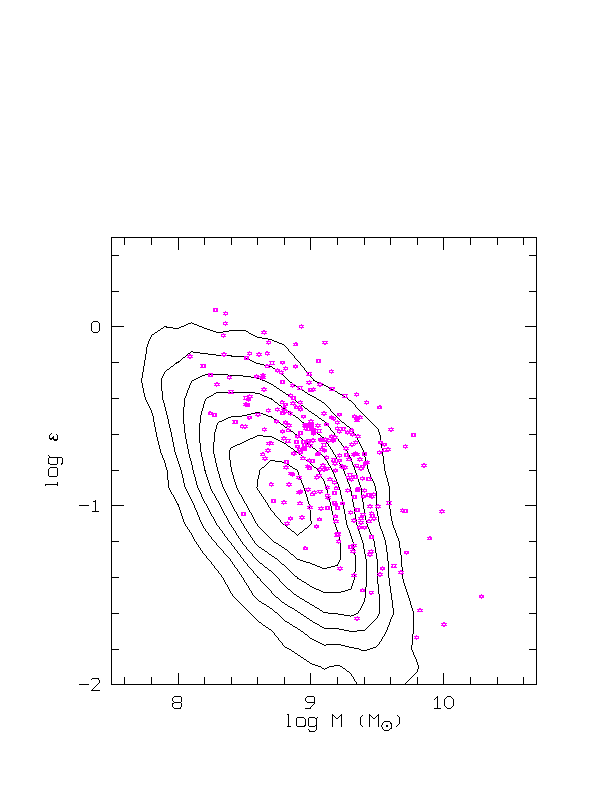}\hfill
\includegraphics[viewport=0 52 550 560,angle=0,width=6.0cm,clip]{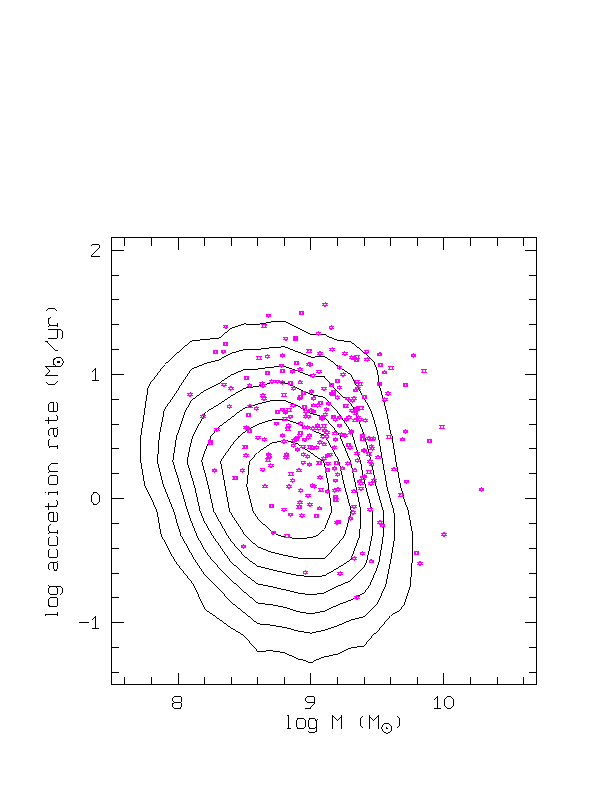}\hfill
\end{tabbing}
\caption{Bolometric luminosity (left), Eddington ratio $\varepsilon$ (middle), and accretion rate (right)
as a function the black hole mass. 
Top: rWLQ sample (magenta frames: rWLQ-EWS subsample, filled red squares: FIRST-detected, open red squares: 
FIRST undetected) and comparison sample (CS; contours).
Second row: high-luminosity subsample (blue symbols) from a simulated 
quasar sample (contours).
Third row: 
The subsample of comparison quasars with $W_\ion{Mg}{ii}<15$\AA\ and $W_\ion{C}{iv}<7$\AA\ (black symbols)
compared with the entire comparison sample (contours).
Bottom: 
quasars with high-S/N spectra (magenta symbols) from the comparison sample 
compared with the entire comparison sample (contours).
}
\label{fig:all_parameters}
\end{figure*}

Table\,\ref{tab:mean-prop} suggests that our WLQs tend to have on average higher luminosities, black hole masses, 
Eddington ratios, and accretion rates compared to ordinary quasars.  We
tested the null hypothesis $H0$ that the WLQ samples and the comparison sample are drawn from the same
population against the alternative hypothesis $H1$ that the values of the population from which the WLQs were 
drawn are statistically higher than the values of the population of ordinary quasars. We applied the 
one-tailed Kolmogorov-Smirnov two sample test (e.g. Siegel \& Castellan \cite{Siegel1988}). 
$H0$ has to be rejected with an error probability $\alpha$ if the 
test statistic $D_{\rm max}$ is larger than a critical value $D_{\rm crit, \alpha} (n_1,n_2)$, where $\alpha$ is the 
probability of rejecting $H0$ when it is in fact true, $n_1$ and $n_2$ are the numbers of quasars in the two samples.
Table\,\ref{tab:KS-test} lists the values for $D_{\rm max}$ for 
the first seven samples from Tab.\,\ref{tab:mean-prop}, the subsample of lobe-dominant radio-loud WLQ was omitted 
because of the small sample size. In the last two columns, the critical values $D_{\rm crit}$ are listed  
for $\alpha$ = 0.01 and 0.001, respectively. With two exceptions, $H0$ has to be rejected in favour of $H1$ at
$\alpha = 0.001$.  The null hypothesis cannot be rejected for the black hole masses of the EWS subsamples. 
There is no strong difference, on the other hand, between WLQs of different radio properties, although the 
radio-loud subsample is a little bit more extreme than  the radio-quiet one.

Figure\,\ref{fig:all_parameters} displays the distributions in the $L-M-\varepsilon-\dot{M}$ parameter space.
The top row shows the rWLQ-EWS sample and the rWLQ sample as symbols and the comparison sample by contours. 
The WLQ sample contains several extremely luminous quasars. The highest luminosity is measured 
for SDSS J152156.48+520238.5 with $\log L = 48.1$, which is one of the four most 
luminous quasars in the SDSS DR7 quasar catalogue.\footnote{The luminosity $\log L = 48.29$ of the 
most luminous quasar, SDSS J074521.78+473436.1, is only marginally higher.} 
Also the black hole mass $M = 1.3\,10^{10}\,M_\odot$ and the Eddington ratio $\varepsilon = 0.95$ of 
SDSS J152156.48+520238.5 are very high,  whereas the accretion rate $\dot{M} = 2.1 M_\odot/$yr is 
rather normal. 
However, these values should be interpreted with caution because the virial mass is derived from the
\ion{C}{iv} line (see Sect. \ref{sec:selection}).
High-quality data provided by Wu et al. (\cite{Wu2011}), based on near-infrared spectroscopy, yield 
$M = 6.2\,10^{9}\,M_\odot$ and $\varepsilon = 0.81$. With $z = 2.238$ (Wu et al. \cite{Wu2011}),
this quasar does not belong to the 
rWLQ sample, but there is also an overabundance of very luminous quasars in the rWLQ sample.  
On the other hand, the top left panel of Fig.\,\ref{fig:all_parameters} indicates a lower luminosity threshold 
for the rWLQs at $\log L_{\rm low} \approx 46.3$, though there is some scatter.  
The higher mass of the rWLQs is probably the consequence of the correlation between $L$ and $M$ in the
parent sample (see also Meusinger \& Weiss \cite{Meusinger2013}). 
There is no significant difference between the distributions of the radio-loud subsample and the rest, though
radio-loud WLQs tend to be slightly more luminous.

\begin{figure}[bhtp] 
\includegraphics[viewport=30 30 590 800,angle=270,width=9.2cm,clip]{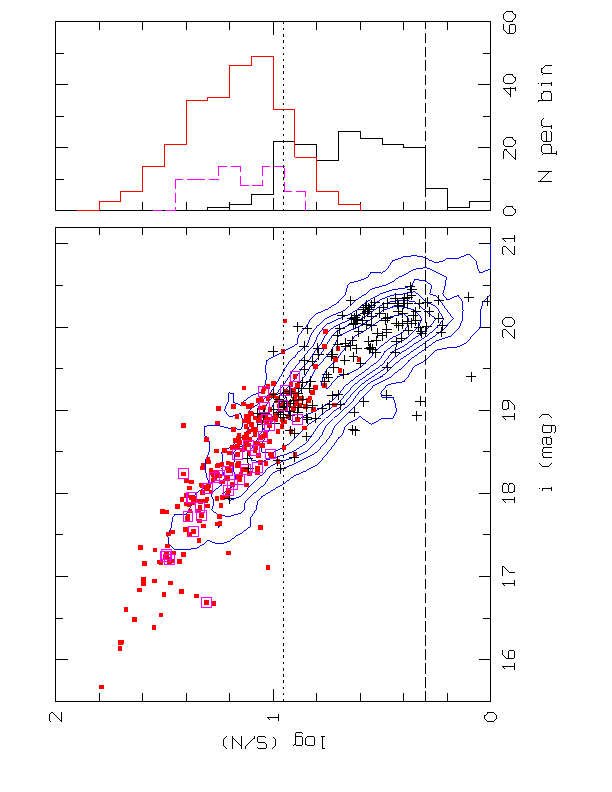} 
\caption{Left: Signal-to-noise ratio versus i band magnitude for the rWLQ sample (filled red squares; 
magenta frames: rWLQ-EWS subsample) and for the WLQ candidates that were rejected as noisy 
(black plus signs). For comparison, the contour curves show the distribution of the comparison sample 
(equally spaced logarithmic local point density contours estimated with a 
grid size of $\Delta x,\Delta y = 0.1,0.05$). Horizontal dashed line: explicit selection 
threshold S/N$>2$, dotted line: 12th percentile of the rWLQ sample and 88th percentile of the
unclassified noisy spectra at S/N=9.
Right: Frequency distribution for the rWLQ sample (solid red), the rWLQ-EWS subsample
(dashed magenta), and the WLQ candidates rejected as noisy (solid black). The histogram for the 
rWLQ-EWS subsample is stretched by a factor of two for better visibility.
}
\label{fig:snr_i}
\end{figure}

\begin{figure*}[bhtp]
\includegraphics[viewport=0 0 550 780,angle=0,width=5.15cm,clip]{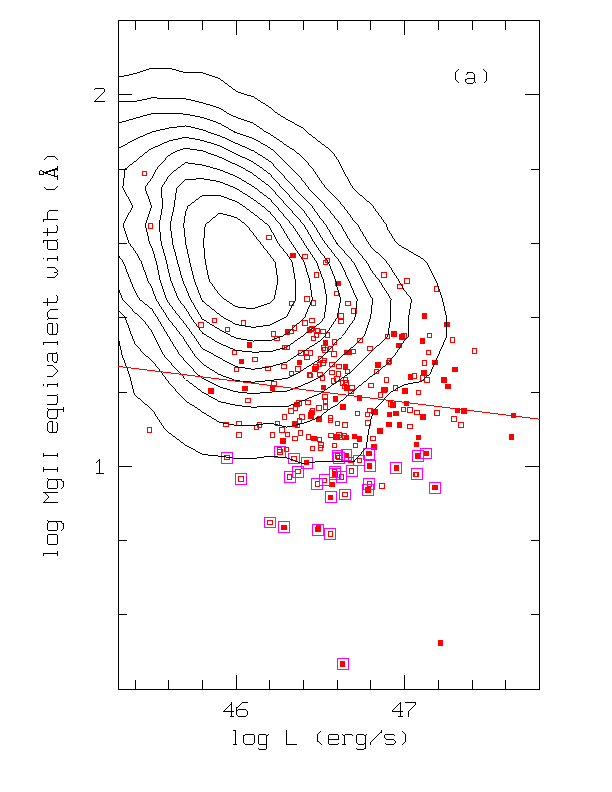}
\includegraphics[viewport=100 0 550 780,angle=0,width=4.2cm,clip]{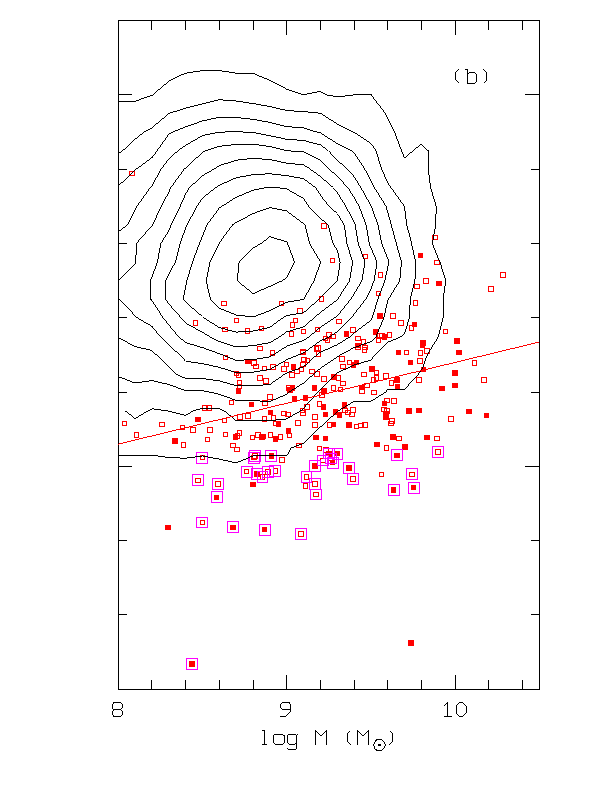}
\includegraphics[viewport=110 0 550 780,angle=0,width=4.12cm,clip]{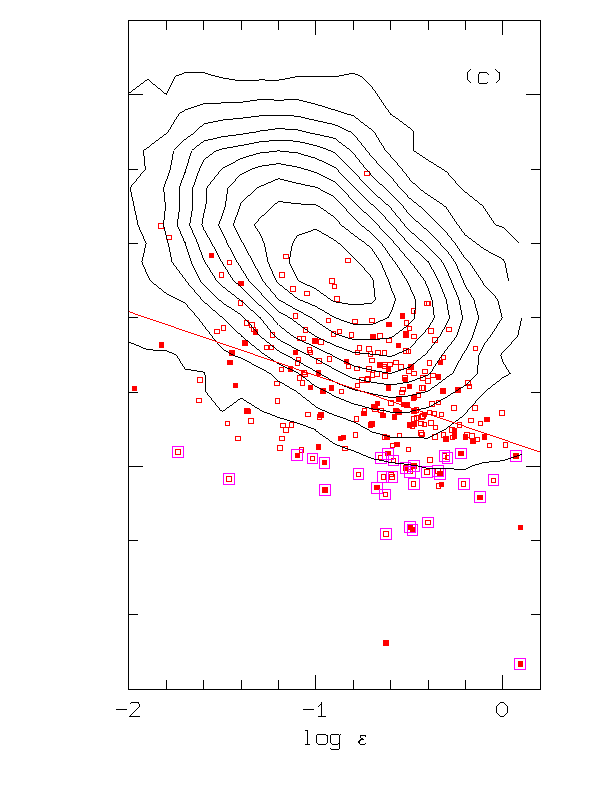}
\includegraphics[viewport=100 0 550 780,angle=0,width=4.2cm,clip]{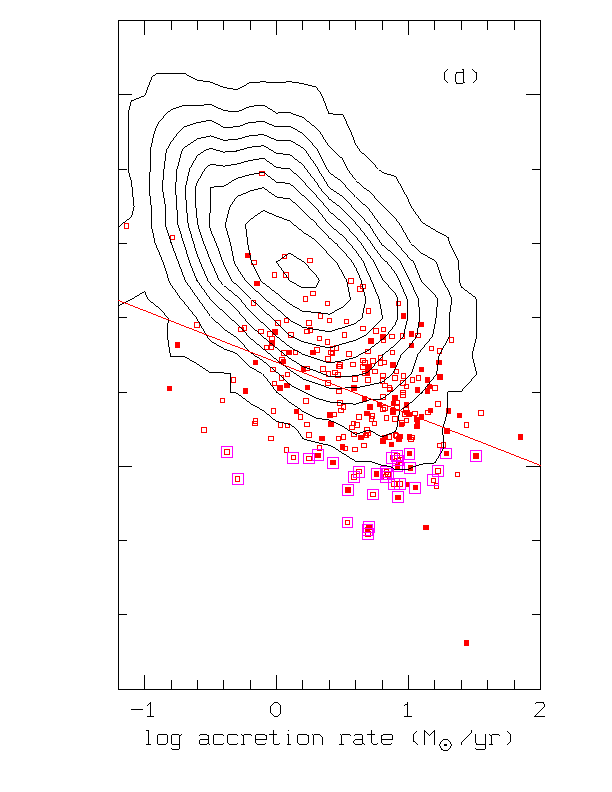}
\caption{Equivalent width of the \ion{Mg}{ii} line versus (a) bolometric luminosity,
(b) black hole mass, (c) Eddington ratio, and (d) accretion rate for the rWLQ sample.
Filled red squares: FIRST detected WLQs, open red squares: FIRST undetected,
magenta frames: rWLQ-EWS subsample.
Straight lines: linear regression curves for the rWLQ sample.
The distributions for the comparison sample are indicated by equally
spaced logarithmic local point density contours estimated with a grid size of
$\Delta x,\Delta y = 0.1,0.05$ dex.}
\label{fig:Edd_ratio}
\end{figure*}

Because the lower luminosity cut does obviously not significantly depend on $M$, it produces $M$-dependent 
lower limits of the Eddington ratio with $\log \varepsilon_{\rm low} \propto -\log M$ and of the accretion rate
with $\log \dot{M}_{\rm low} \propto -0.89 \log M$ that are clearly seen in the corresponding panels in the top
row of Fig.\,\ref{fig:all_parameters}.  We demonstrate such an effect by a simple simulation. 
The approach is described in detail in Meusinger \& Weiss (\cite{Meusinger2013}), 
here we only repeat the most important steps. 
We simulated a quasar sample with a $z$ distribution similar to that of the rWLQ sample and with a
reasonable mass distribution. Then we assigned a randomly chosen value of the radiative efficiency 
of the accretion process $\eta = 0.057\ldots 0.321$ for non-rotating to maximum spin black holes. 
With $\eta = L/(\dot{M}c^2)$ and the scaling relation for $\dot{M}$, we get a relation between $\eta, L,$ 
and $M$ and can thus compute $L$ for each quasar 
assuming, for simplicity, the same spectral slope $\alpha_\lambda = -1.52$ for all quasars.
We applied a $z$-dependent lower $L$ limit 
corresponding to the flux limit of the survey, an upper $L$ limit corresponding to the luminosity 
distribution, and Gaussian distributed errors of the ``observed'' mass. This simulated ``observed''
quasar sample shows distributions on the $L - M$, $\varepsilon - M$, and $\dot{M}-M$ planes  
similar to those of our comparison sample of ordinary quasars. 
The differences may be due primarily to the very rough assumptions on the mass distribution and the $\eta-M$ relation 
and are not relevant here. Next, we created a subsample of high-luminosity quasars simply by 
arbitrarily selecting a proportion (20 \%) of the quasars with $\log L > 46.3$. The simulated sample and
the high-$L$ subsample are shown in the second row of Fig.\,\ref{fig:all_parameters}.
The distributions in the $\varepsilon-M$ plane and in the $\dot{M}-M$ plane are very similar
to the observed ones in the top row of Fig.\,\ref{fig:all_parameters}. Such an agreement cannot be achieved for
a subsample determined by a threshold of the Eddington ratio. Therefore, we conclude that
the visually selected WLQ sample is better characterised by higher luminosities. The differences 
in $\varepsilon$ and $\dot{M}$ are then at least partly a consequence of the higher $L$ in combination 
with the intrinsic $L-M$ relation of the parent SDSS quasar sample and the $\varepsilon-L$ or $\dot{M}-M$ relation, 
respectively.

The third row of Fig.\,\ref{fig:all_parameters} shows a subsample of low-EW quasars drawn from the
comparison sample of ordinary quasars by an EW threshold. There are too few quasars in the comparison
sample that satisfy the selection criterion of the rWLQ-EWS subsample. Therefore, the constraint
had to be relaxed to $W_\ion{Mg}{ii}<15$\,\AA\ and $W_\ion{C}{iv}<7$\,\AA. The thus selected subsample
appears to be characterised by higher Eddington ratios, accretion rates, and luminosities compared
with the parent sample, but clearly not by higher masses.

Finally, the bottom row of Fig.\,\ref{fig:all_parameters} shows another subsample from the
comparison sample. Here, only quasars with high S/N of their spectra, S/N $>9$, were selected.
The distributions are remarkably similar to those of the rWLQ sample in the top row.

\subsubsection{S/N bias}

Is the tendency of WLQs to be more luminous than ordinary quasars at the same $z$ a selection effect? 
When measuring the spectral index $\alpha_\lambda$, the mean S/N in the pseudo continuum windows was 
estimated and quasars with a mean S/N $<2$ were excluded. However, the number of the rejected quasars is very 
low and this explicit selection threshold is thus not very important. A much stronger 
bias towards higher S/N might be implicitly introduced by the visual selection from the icon maps. 
A certain classification as a WLQ candidate requires sufficient S/N and S/N is tightly correlated
with the flux density.

Figure\,\ref{fig:snr_i} shows the measured S/N versus SDSS i band magnitude for both the rWLQ sample 
and the comparison sample. The distribution of the rWLQ-EWS subsample is similar to that of the rWLQ sample.
The rWLQ sample only populates the upper part of the region populated by the comparison sample. There are two
possible explanations for this difference: WLQs either have higher S/N due to a selection effect 
or are more luminous than normal quasars. In reality, both effects can be combined, of course.
When we consider the distribution of those quasars that were rejected in the course 
of the classification procedure because the spectra appeared too noisy, we find a 
dichotomy. 88\% of the rWLQ sample have S/N = 9, while about the same percentage of the 
objects with noisy spectra are below that value. Taking S/N = 9 as representative for
the selection threshold and applying this threshold to the comparison sample, we find the mean
luminosity of the high-S/N subsample to be about two times higher than that of the entire comparison 
sample.  The quasars from the rWLQ sample are on average $\sim 4$ times brighter than normal
(Tab.\,\ref{tab:mean-prop}).  About half of the luminosity excess is explained by the S/N 
selection bias.

To summarise, Figs.\,\ref{fig:all_parameters} and \ref{fig:snr_i} along with Tab.\,\ref{tab:mean-prop}
suggest that WLQs have higher luminosities, Eddington ratios, and accretion rates than normal quasars, although 
our basic sample is strongly affected by a selection bias. The relatively high masses of our WLQs are most
likely a consequence of this selection effect.

\begin{figure*}[bhtp]
\begin{tabbing}
\includegraphics[viewport=15 35 550 800,angle=0,width=4.5cm,clip]{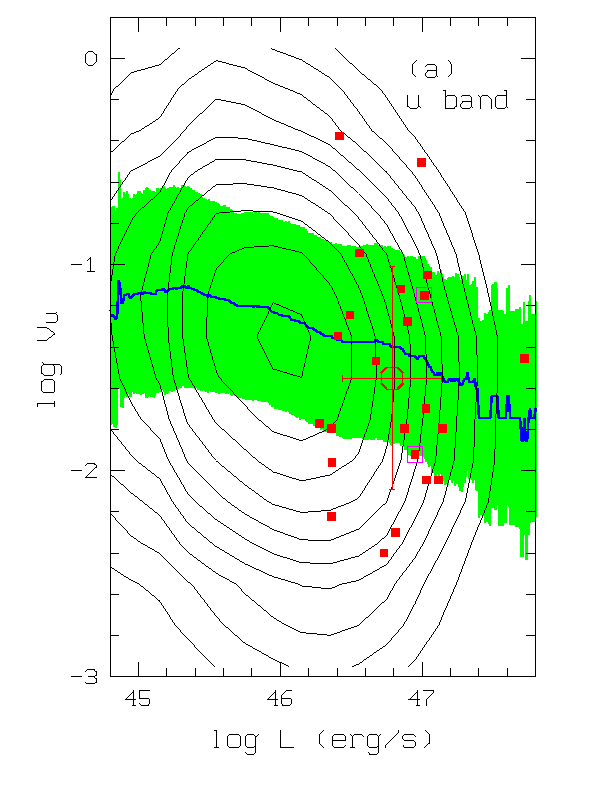}
\includegraphics[viewport=15 35 550 800,angle=0,width=4.5cm,clip]{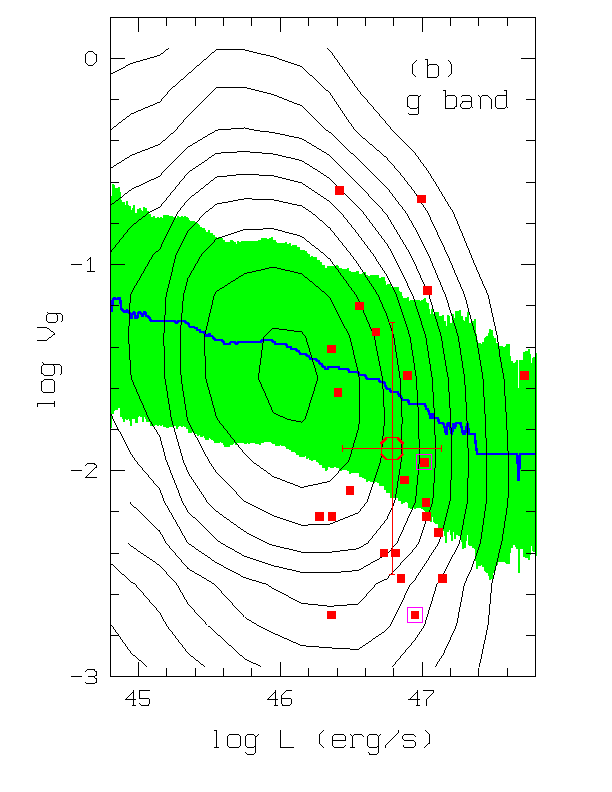}
\includegraphics[viewport=15 35 550 800,angle=0,width=4.5cm,clip]{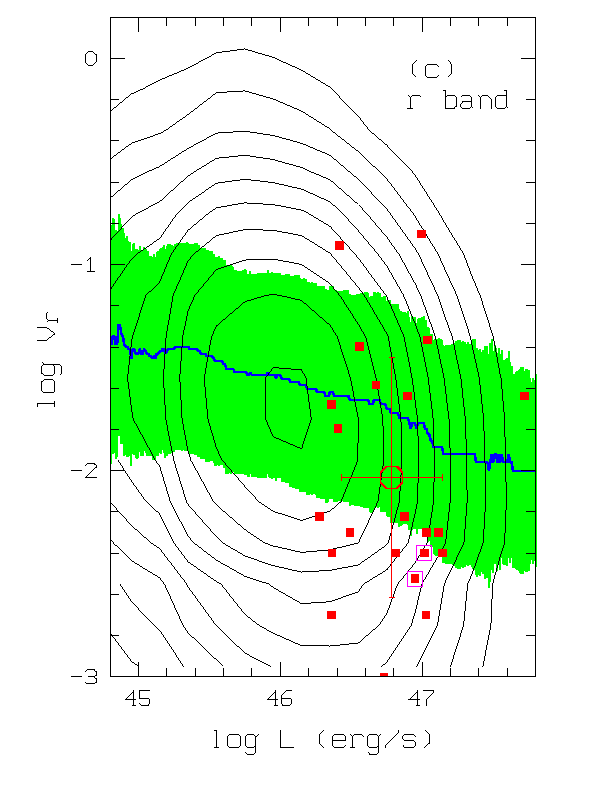}
\includegraphics[viewport=15 35 550 800,angle=0,width=4.5cm,clip]{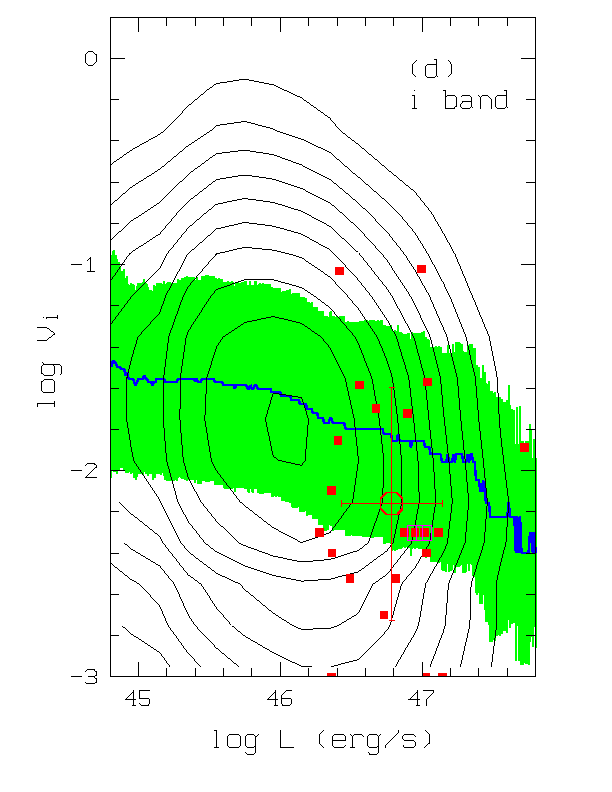}
\end{tabbing}
\caption{
Variability index $V$ in the u, g, r, and i band (left to right) versus luminosity for the 23 WLQs
identified in the variability catalogue (filled red squares, magenta frames:
WLQ-EWS subsample).
For comparison, the entire sample from the variability catalogue
is indicated by the median relation (thick blue curve) with standard deviation (shaded green area)
and by the equally spaced logarithmic local point density contours (estimated with a grid size of
$\Delta x,\Delta y = 0.2,0.2$). Open circle with error bars:
mean value and 1$\sigma$ deviations of the WLQs.
}
\label{fig:var}
\end{figure*}

\subsubsection{Baldwin effect}

The anti-correlation between the luminosity\footnote{More precisely, the rest-frame continuum luminosity at
1450\,\AA.} and the EW of the \ion{C}{iv} line was first discovered by Baldwin (\cite{Baldwin1977}) and has 
become known as the Baldwin effect. A number of subsequent studies have confirmed this effect for the 
\ion{C}{iv} line and for many other emission lines in the ultraviolet and optical
(e.g. Green et al. \cite{Green2001};
Dietrich et al. \cite{Dietrich2002};
Wu et al. \cite{Wu2009};
Richards et al. \cite{Richards2011};
Bian et al. \cite{Bian2012}).
Several interpretations about the origin of the Baldwin effect have been proposed, but there 
is currently no consensus about the fundamental mechanisms behind it. A promising explanation
is that it is related to a luminosity-dependence of the SED shape  
(e.g. Netzer et al. \cite{Netzer1992};
Zheng \& Malkan \cite{Zheng1993};
Dietrich et al. \cite{Dietrich2002};
Wu et al. \cite{Wu2009}). In the last decade, several studies proposed that the 
Baldwin effect may be driven by the Eddington ratio
(Shang et al. \cite{Shang2003}; 
Bachev et al. \cite{Bachev2004}; 
Baskin \& Laor \cite{Baskin2004};
Dong et al. \cite{Dong2009, Dong2011}; 
Bian et al. \cite{Bian2012}).

In Fig.\,\ref{fig:Edd_ratio}, the \ion{Mg}{ii} equivalent width is plotted versus $L, M, \varepsilon,$ and $\dot{M}$, 
respectively. The Baldwin effect is clearly indicated for the comparison sample where we observe
slightly stronger anti-correlations  with $\varepsilon$ and $\dot{M}$  than with $L$.
The strongest effect is found for $\dot{M}$ probably indicationg that the accretion rate is the main driver.
No significant correlation is indicated with the black hole mass $M$. 
As in Fig.\,\ref{fig:all_parameters}, there is no significant difference between 
radio-loud and not radio-loud quasars. Similar to Fig.\,\ref{fig:all_parameters},
the areas populated by the rWLQ sample  in Fig.\,\ref{fig:Edd_ratio} 
are identical with the regions of low $W$ and high $L$, $\varepsilon$, and $\dot{M}$ of normal quasars.
The rWLQ-EWS sample is simply the low-EW part
of the rWLQ sample. (Note that there are a few WLQs with $W_\ion{Mg}{ii}$ below but 
$W_\ion{C}{iv}$ above the EWS threshold.) The high mean values of the luminosity, the Eddington ratios,
and the accretion rate in Tab.\,\ref{tab:mean-prop} are thus partly a reflex of the Baldwin effect.
The Baldwin effect is also present in the rWLQ sample 
(regression lines). Because of the inherent selection bias it is, however, not useful 
to compare the slopes with those from the literature

\subsection{Variability}\label{subsec:variability}

In a previous study (Meusinger et al. \cite{Meusinger2011}), we exploited the 
Light-Motion Curve Catalogue (LMCC; Bramich et al. \cite{Bramich08})
of 3.7 million objects with multi-epoch photometry from the S82 of the   
SDSS DR7 to analyse the light curves for about 9\,000 quasars in the five SDSS bands. 
We computed the rest-frame first-order structure function (SF)  $D(\tau_{\rm rf})$ for
each LMCC light curve in each band.
The SF is a sort of running variance of the magnitudes as a function
of the rest-frame time-lag $\tau_{\rm rf}$, i.e. the time difference between two
arbitrary measurements. The arithmetic mean of all noise-corrected SF data points in the interval
$\tau_{\rm rf, max} \approx 1-2$\,yr was taken as variability estimator. In other words, the quantity 
$V$ used to describe the strength of the variability of a quasar is the variance of its magnitude 
differences from all those pairs of two measurements that have rest-frame time-lags between 1 and 2 yr.
$V$ was computed for each of the five SDSS bands.

The sample from our previous study included three quasars from the WLQ sample of Diamond-Stanic et al. 
(\cite{Diamond2009}). Variability was found to be close to the median value for one of them
but is remarkably lower for the other two. 
Also other quasars with relatively weak BELs selected from the same sample 
were found to vary by less than the general median dispersion. However, this previous WLQ sample was 
very low and the interpretation of some of these objects remains uncertain.

We identified 23 WLQs from our present sample with entries in the variability catalogue
of S82 quasars. Using the variability estimators from this catalogue, we found again a tendency 
towards lower variability for the WLQs compared to the entire sample.
Figure\,\ref{fig:var} shows the variability-luminosity diagrams using the variability indices 
$V$ in the u, g, r, and i band, respectively.

It has been known for a long time that variability is anti-correlated with luminosity, in the sense 
that, at a given rest-frame wavelength, more luminous AGNs tend to vary with lower fractional 
amplitudes than less luminous AGNs (e.g.
Pica \& Smith \cite{Pica1983}; 
Hook et al. \cite{Hook1994};
Paltani \& Courvoisier \cite{Paltani1997};
Vanden Berk et al. \cite{VandenBerk2004};
De Vries et al. \cite{DeVries2005};
Meusinger et al. \cite{Meusinger2011}).
The  main driver behind the $V-L$ relation may be the Eddington ratio or the accretion rate
(e.g. Ai et al. \cite{Ai2010}; Meusinger \& Weiss \cite{Meusinger2013}).
A tendency towards lower variability indices of our WLQs is thus expected as a consequence of their higher 
luminosities. However, the mean values for the 23 WLQs are below the mean $V-L$ relation in all SDSS bands. 
Figure\,\ref{fig:var} thus indicates that the continuum in WLQs is, on average, at least not more variable than 
in ordinary quasars. This conclusion is boosted by the fact that the variability of the line flux is weaker 
than the variability of the underlying continuum (Wilhite et al. \cite{Wilhite2005}; 
Meusinger et al. \cite{Meusinger2011}).

%
%
\section{Wide band spectral energy distribution}\label{sec:SED}
%
%

\begin{figure*}[bhtp]
\begin{tabbing}
\includegraphics[viewport=10 100 550 670,angle=0,width=4.6cm,clip]{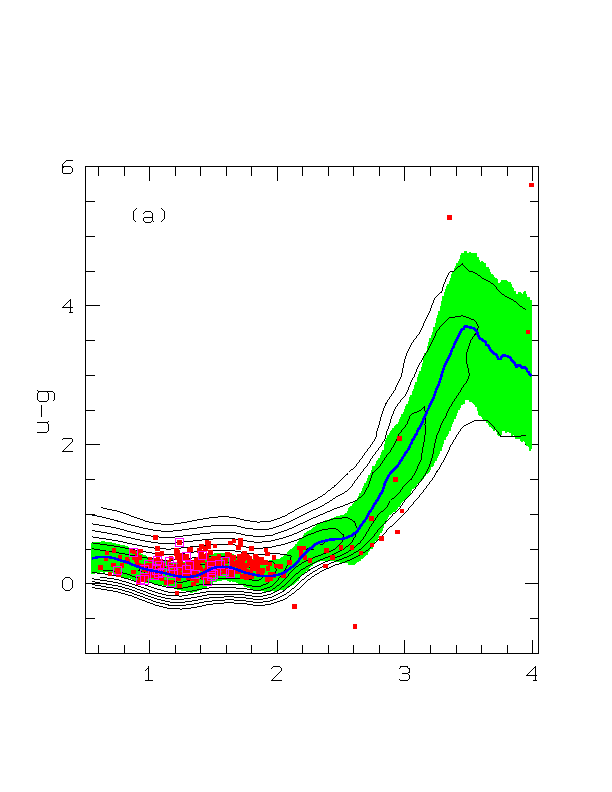}\hfill
\includegraphics[viewport=10 100 550 670,angle=0,width=4.6cm,clip]{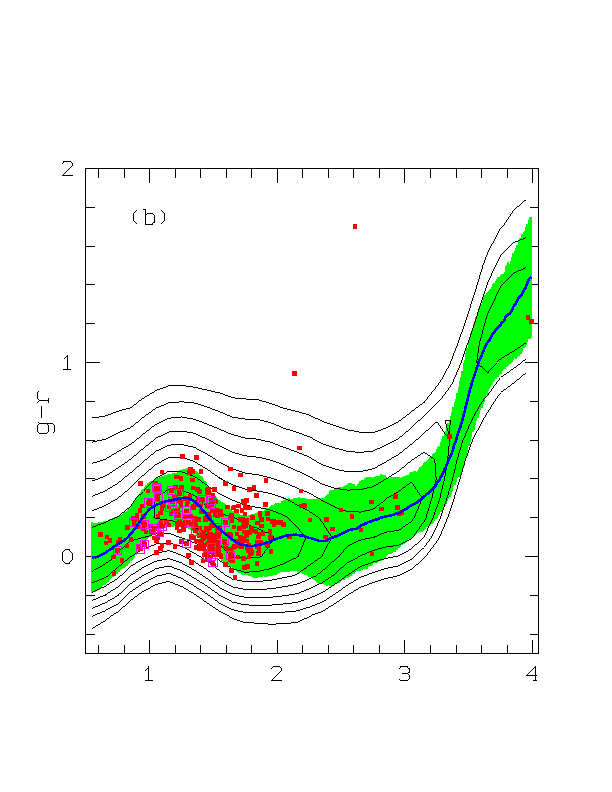}\hfill
\includegraphics[viewport=10 100 550 670,angle=0,width=4.6cm,clip]{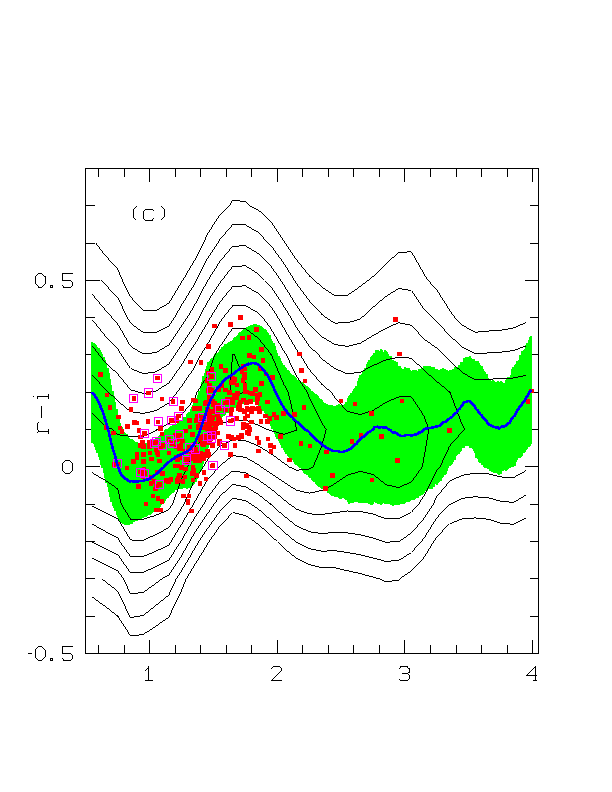}\hfill
\includegraphics[viewport=10 100 550 670,angle=0,width=4.6cm,clip]{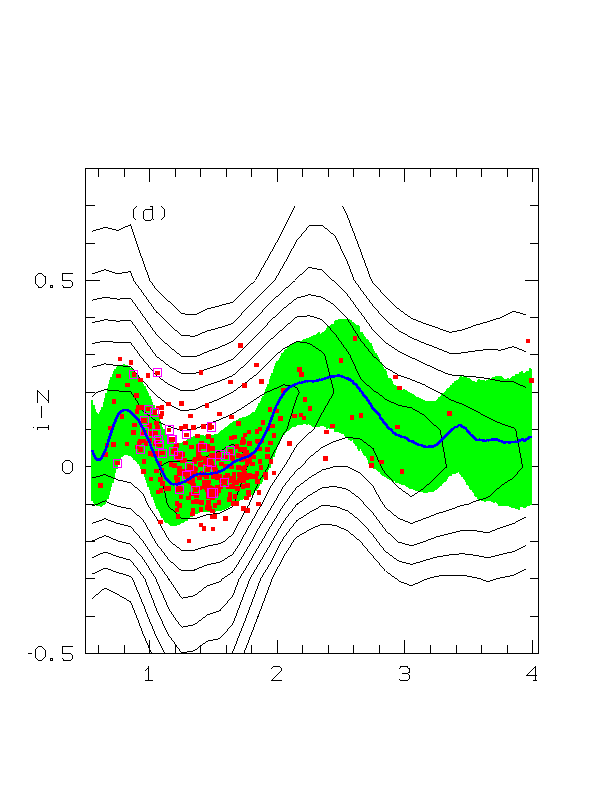}\hfill \\
\includegraphics[viewport=10 60 550 650,angle=0,width=4.6cm,clip]{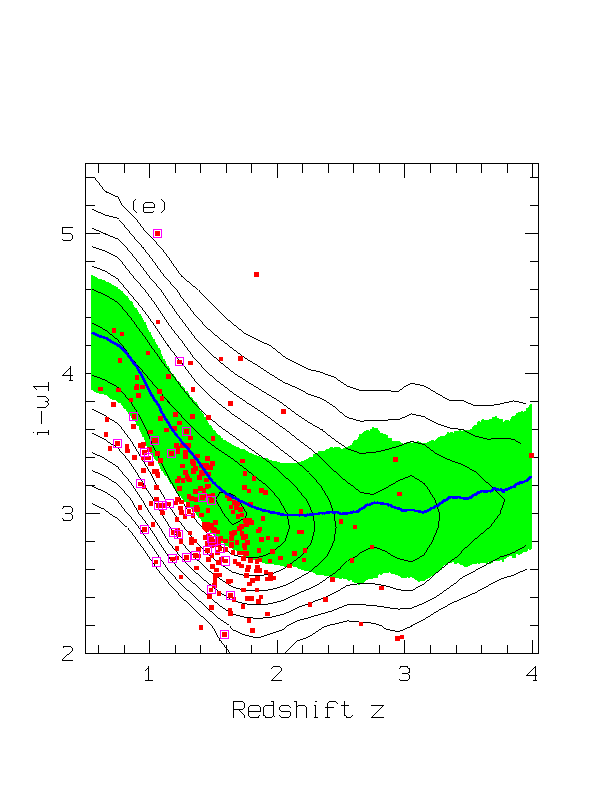}\hfill
\includegraphics[viewport=10 60 550 650,angle=0,width=4.6cm,clip]{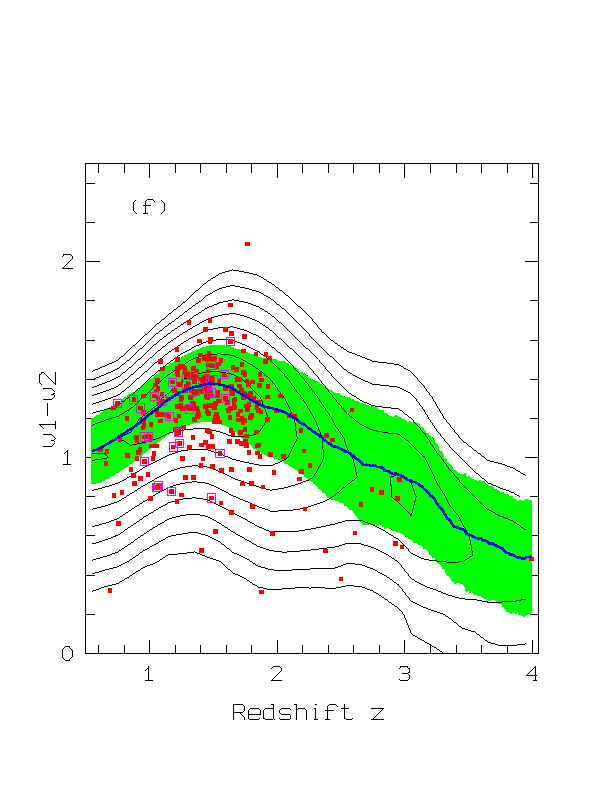}\hfill
\includegraphics[viewport=10 60 550 650,angle=0,width=4.6cm,clip]{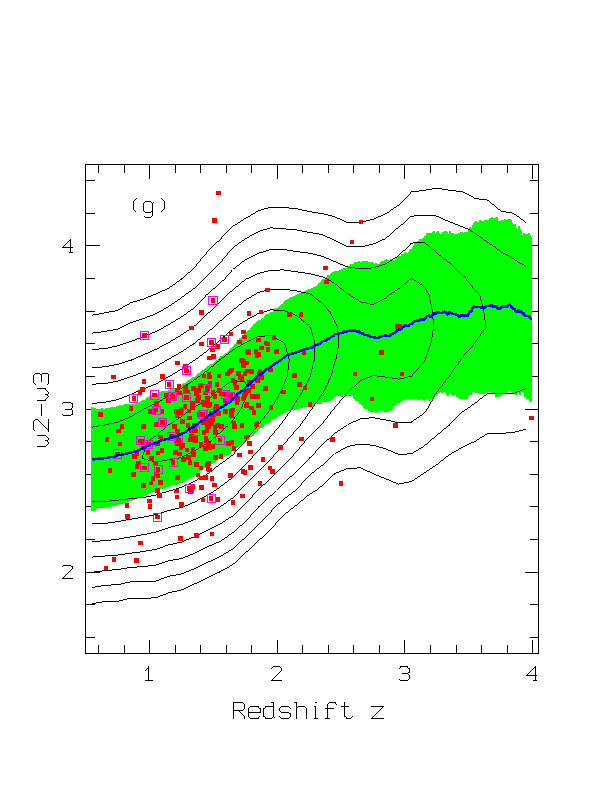}\hfill
\includegraphics[viewport=10 60 550 650,angle=0,width=4.6cm,clip]{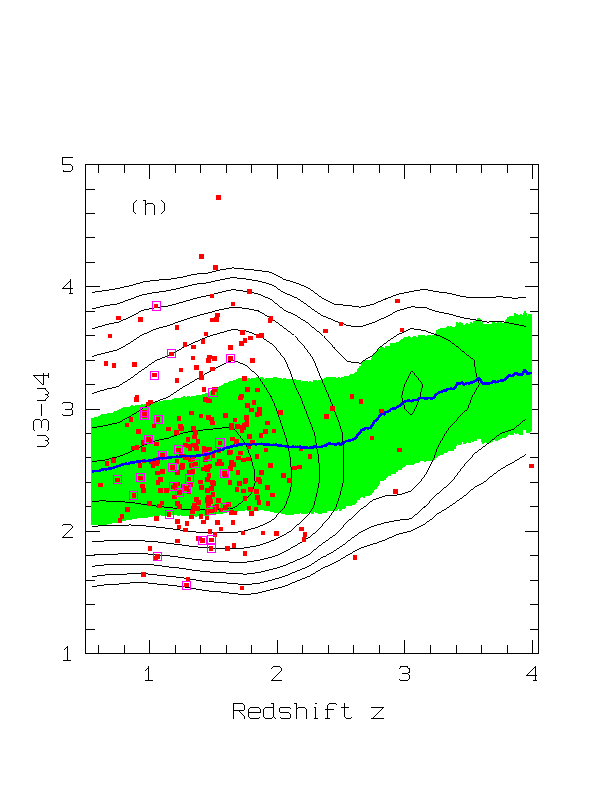}\hfill
\end{tabbing}                                                          
\caption{SDSS and WISE colour-redshift diagrams for the WLQ sample (filled red squares;
magenta frames: WLQ-EWS subsample).
For comparison, the distributions of all quasars from the Shen catalogue are shown by contours
(equally spaced logarithmic local point density contours estimated with a grid size of
$\Delta x,\Delta y = 0.1,0.1$) and by the median
colour redshift relation (thick blue curve) with standard deviation (shaded green area).
}
\label{fig:SDSS_WISE}
\end{figure*}

\subsection{From Ly$\alpha$ to the mid-infrared}\label{subsec:wide-band}

Information on the SED of the quasars with a much wider wavelength coverage than 
the SDSS spectra is available particularly thanks to the sky surveys in the infrared.
We used the photometric data for the J, H, and K band from the Two Micron All-Sky Survey (2MASS; Skrutskie et al. 
\cite{Skrutskie2006}) and for the w1 to w4 bands from the Wide-Field Infrared Survey Explorer 
(WISE; Wright et al. \cite{Wright2010}) in combination with the u,g,r,i, and z magnitudes from the SDSS. 
The SDSS and  2MASS magnitudes were taken from the SDSS DR7 quasar catalogue (Schneider et al. \cite{Schneider2010}). 
We identified 95\% of the SDSS DR7 quasars in the WISE All-Sky Source 
Catalog\footnote{http://irsa.ipac.caltech.edu/cgi-bin/Gator/nph-dd} within a search radius of $6\arcsec$. 
99\% of the identified quasars have SDSS-WISE position differences less than $3\arcsec$. 
2MASS J, H, and K band magnitudes are only available for 38, 28, and 28\%, respectively.

The colour-redshift diagrams from the SDSS and WISE data are shown in Fig.\,\ref{fig:SDSS_WISE}. 
Although at first glance the mean colours of the WLQs are similar to those of the entire quasar 
population, the wavy structure is less pronounced for the WLQs and
even less for the WLQ-EWS subsample.
The features in the SDSS colour-$z$ diagrams can be understood as being caused by the strong emission lines 
moving in and out of the filters with changing redshift (e.g. Richards et al. \cite{Richards2001}).
For example, normal quasars have a relatively blue $r-i$ at $z \approx 1$, where the \ion{Mg}{ii} line
dominates the r band, and a relatively red $r-i$ at $z \approx 1.7$ where it falls into the i band.
The weaker \ion{Mg}{ii} line of the WLQs results in a redder colour $r-i$ at
$z \approx 1$ and a bluer at  $z \approx 1.7$ compared to the median colour redshift relation.
There are no systematic differences 
in the WISE colour-$z$ diagrams between the WLQs and the normal quasars. 
For the rWLQ sample, we derived 
mean colour indices
$w1-w2 = 1.22\pm0.25, w2-w3 = 2.97\pm0.33, w3-w4 = 2.66\pm0.52$ compared to 
$w1-w2 = 1.28\pm0.22, w2-w3 = 2.97\pm0.35, w3-w4 = 2.71\pm0.51$ for the comparison samples.
However, in a colour-$z$ diagram for a combined  SDSS-WISE colour, the WLQs appear to be significantly 
bluer on average, as shown for $i-w1$ in the Fig.\,\ref{fig:SDSS_WISE}\,e. 
For  $z=0.7\ldots 2$, this colour index reflects the slope of the SED between rest-frame near-ultraviolet 
and near-infrared.

After applying Galactic extinction corrections to the photometry in all 12 bands, the magnitudes were transformed 
into monochromatic fluxes per wavelength interval, $F_\lambda$, and the effective wavelengths of each filter
were transformed into rest-frame wavelengths. 
We arbitrarily normalised the rest-frame SED of each quasar at 3000 \AA. Therefore, the 
redshift range had to be restricted to $ 0.3 \le z \le 2$. The sample contains 79\,055 quasars.
After normalisation, we computed the arithmetic median and 
standard deviation of all data points in narrow wavelength intervals.

\begin{figure*}[bhtp]
\begin{tabbing}
\includegraphics[viewport=40 50 570 780,angle=270,width=9.1cm,clip]{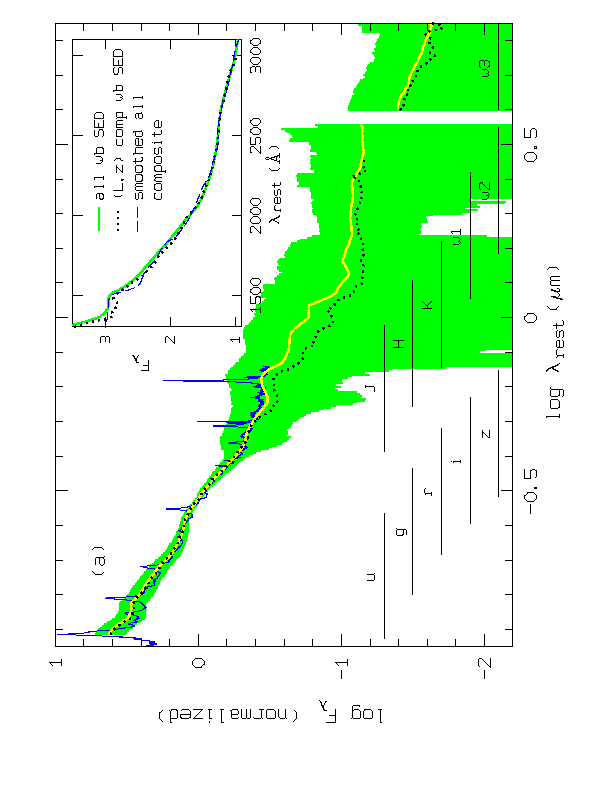}
\includegraphics[viewport=40 50 570 780,angle=270,width=9.1cm,clip]{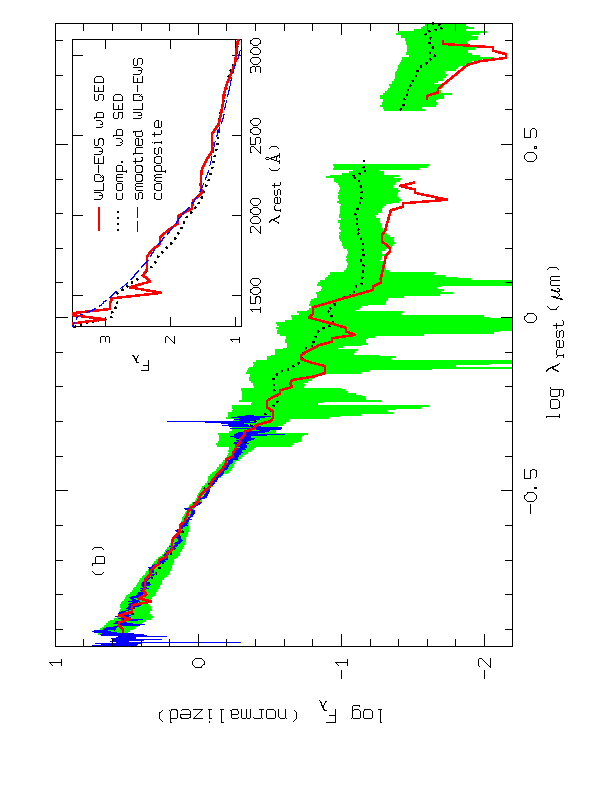}\\
\includegraphics[viewport=40 50 570 780,angle=270,width=9.1cm,clip]{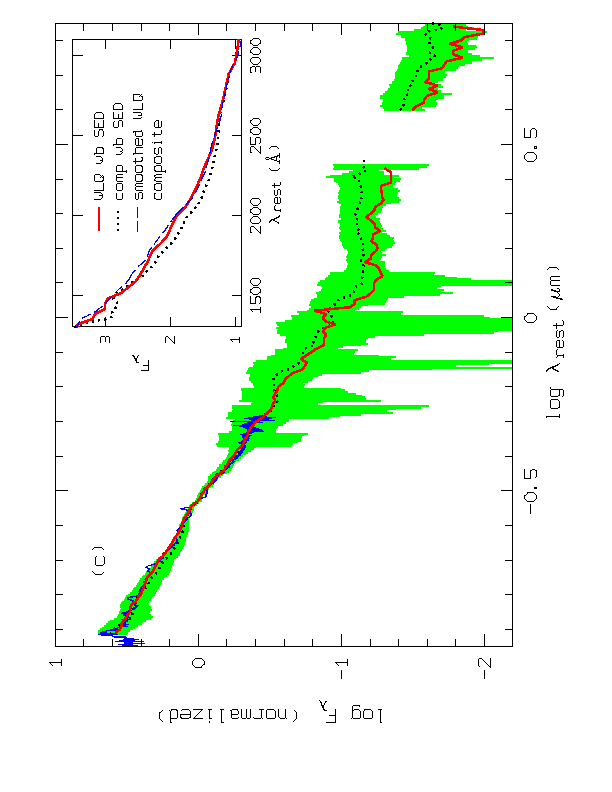}
\includegraphics[viewport=40 50 570 780,angle=270,width=9.1cm,clip]{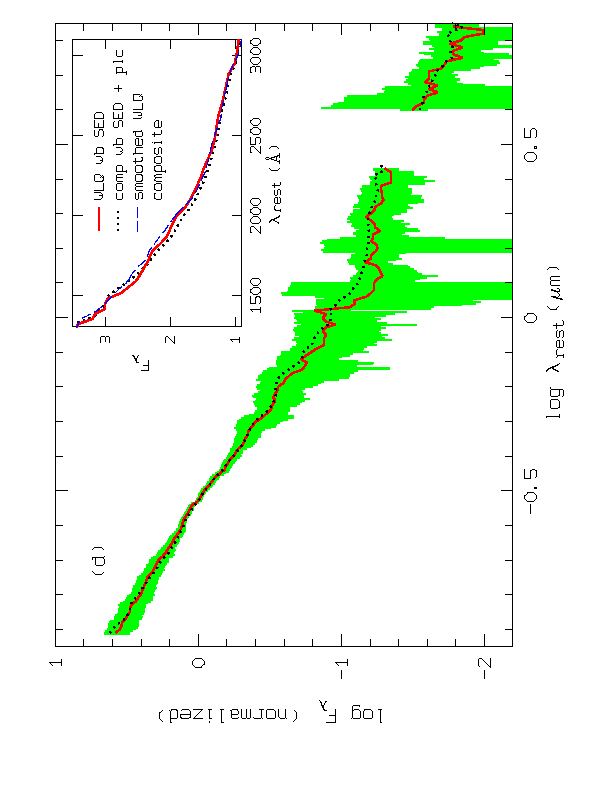} 
\end{tabbing}
\caption{
Wide band (wb) SED. 
(a) Shen catalogue with $z<2$ (yellow solid curve: median, green shaded area: $1\sigma$ errors)
and WLQ comparison sample (black dotted curve), normalised at 3000 \AA. Over-plotted (blue): SDSS quasar
composite spectrum from Vanden Berk et al. (\cite{VandenBerk2001}). 
The horizontal lines indicate the wavelengths intervals covered by the different photometric bands.
The inset shows the wb SED of all SDSS quasars (solid green) and of the comparison sample (black dotted) in the 
wavelength range 1300\AA\ $ - 3000$\,\AA\ in linear scale, as well as the SDSS quasar composite spectrum
smoothed with a 300 \AA\ boxcar (blue dashed).
(b) WLQ-EWS sample (red solid curve) and comparison sample 
(black dotted curve and green shaded area).
Over-plotted (blue): SDSS WLQ composite spectrum from Fig.\,\ref{fig:composite}.
(c) As (b), but for the entire WLQ sample instead of the WLQ-EWS subsample.
(d) As (b), but the comparison sample SED is boosted by a power-law component (plc) with 
$\alpha_\lambda = -1.7$ and re-normalised, the green shaded area indicates the 1$\sigma$ errors of the WLQ sample,
and the SDSS composite spectrum was omitted for the sake of clarity.
}
\label{fig:wide_band}
\end{figure*}

In Fig.\,\ref{fig:wide_band}\,a, we plotted the arithmetic median wide band SED and the 1$\sigma$ error 
interval for the entire quasar sample from the SDSS DR7 quasar catalogue. It is well matched by the
SDSS composite spectrum from Vanden Berk et al. (\cite{VandenBerk2001}). The de-redshifted wavelengths intervals 
covered by the various photometric bands are indicated by the vertical lines and labelled in the lower part.
It is evident that the structures seen in the 1$\sigma$ area are related to the edges of the bands in combination with
the different photometric errors. In the near infrared, the strong incompleteness of the 2MASS photometry must be mentioned.

Quasar to quasar variations of the host galaxy fraction contribute to the increase of the scatter from UV to optical 
wavelengths. Because the host contamination depends on the quasar luminosity,  the WLQ sample with its higher 
mean luminosity is expected to have a smaller host contribution compared to the entire quasar sample.
We therefore have to compare the SED of the WLQs with that of a comparison sample with the same luminosity distribution. 
We constructed such a control sample in a similar way as mentioned in Sect.\ref{sec:selection}, but this time the 
two-dimensional distribution in the $L-z$ space is required to match that of the WLQ sample. 
The size of this ($L,z$) comparison sample is again ten times that of the WLQ sample.
The wide band SED of the ($L,z$) comparison sample, over-plotted in Fig.\,\ref{fig:wide_band}\,a,
indicates smaller host contributions compared to the parent SDSS quasar sample, as expected.

The wide band SED of the WLQ-EWS sample is shown in  Fig.\,\ref{fig:wide_band}\,b along with  
that of the ($L,z$) comparison sample. The agreement is relatively good, but the WLQ SED turns out to be 
slightly steeper over nearly the covered wavelength range.  The same trend is seen for the 
wide band SED of the entire WLQ sample (Fig.\,\ref{fig:wide_band}\,c) that shows less fluctuations
because of the larger sample size. Figure\,\ref{fig:wide_band}\,d
shows that the agreement between the WLQ sample and the comparison sample is improved
after adding a power-law component
$F_{\rm \lambda,\ add} = c \lambda^{\alpha_\lambda}$ to the comparison sample SED. We checked various parameter
combinations $(c,\alpha_\lambda)$ and found a good match for $\alpha_\lambda \approx -1.8\ldots -1.7$, i.e. 
$\alpha_\nu = -0.2\ldots -0.3$, and an enhancement of the total flux density at 3000 \AA\ by a factor $\sim 2$.

\subsection{Radio emission}\label{subsec:radio}

Table\,\ref{tab:mean-prop} indicates a 
relatively high percentage of radio-loud WLQs compared to the entire SDSS DR7 quasar sample. 
In general, the distribution of the radio-loudness of quasars appears to
be bimodal with about 10\% being radio-loud (Kellermann et al. \cite{Kellermann1989}; White et al. \cite{White2000};
Ivezi\'c et al. \cite{Ivezic2002}).

\begin{table}[htbp]
\begin{center}
\caption{Radio properties of the WLQs and of the comparison samples. 
(\#: number of quasars,
$f_{\rm rl}$: proportion of radio-loud quasars;
$f_{\rm r}$: proportion of radio-detected quasars in FIRST footprint;
$R$: radio-loudness,
$r_{\rm N,cl}$: number ratio of core-dominated to lobe-dominated sources;
$r_{\rm R,cl}$ ratio of mean radio-loudness of core-dominated to lobe-dominated sources)
}
\begin{tabular}{lrccccc}
\hline
Sample         & \# \ \ &  $f_{\rm rl}$  & $f_{\rm r}$ & $\overline{\log R}$  & $r_{\rm N,cl}$ &  $r_{\rm R,cl}$\\
\hline
WLQ-EWS\tablefootmark{a}               &   46  & 0.35  & 0.42 & 1.70 &   8.0 & 0.04  \\
rWLQ-EWS\tablefootmark{a}              &   33  & 0.37  & 0.47 & 1.67 &   6.0 & 0.04  \\
WLQ\tablefootmark{a}                   &  365  & 0.26  & 0.34 & 1.65 &   9.5 & 0.30  \\ 
rWLQ\tablefootmark{a}                  &  261  & 0.22  & 0.33 & 1.58 &   6.9 & 0.22  \\
Comparison\tablefootmark{a}            & 2750  & 0.06  & 0.06 & 2.23 &   2.7 & 0.60  \\
WLQ Paper 1\tablefootmark{b}           &  161  & 0.25  & 0.34 & 1.52 &  54/0 &  -    \\
WLQ Lit\tablefootmark{c}               &   98  & 0.10  & 0.28 & 1.46 &  26.0 & 0.36  \\
WLQ Lit ($z<3$)\tablefootmark{d}       &   25  & 0.24  & 0.36 &    - &   9/0 &  -    \\ 
Shen EW$<15$\tablefootmark{e}          & 1268  & 0.22  & 0.24 & 1.93 &  10.9 & 0.25  \\
\hline
\end{tabular}
\tablefoot{
\tablefoottext{a}{As in Tab.\,\ref{tab:mean-prop}};
\tablefoottext{b}{WLQs from Paper 1 with identification in Shen catalogue};
\tablefoottext{c}{WLQs from literature (see text)};
\tablefoottext{d}{WLQs from literature with $z<3$};
\tablefoottext{e}{EW selected sample from the Shen catalogue in FIRST footprint with $0<W_{\ion{Mg}{ii}}<15$\AA\ 
and {\sc BAL\_FLAG = 0}}.
}
\label{tab:radio-prop}
\end{center}
\end{table}

Table \,\ref{tab:radio-prop} gives an overview of the radio properties of various samples. The 
proportion of radio-loud quasars, i.e. the ratio $f_{\rm rl} = N_{\rm rl}/N_{\rm all}$ of the number 
$N_{\rm rl}$ of radio-loud quasars to the number $N_{\rm all}$ of all WLQs in FIRST footprint amounts 
to 0.26 for the WLQ sample and 0.22 for the rWLQ sample. The radio-loudness proportion is even higher 
for the EWS subsamples (0.35 and 0.37). In the comparison sample, only 6\% of the quasars are radio-loud, 
in good agreement with the value of 8\% found for the SDSS quasar sample (Ivezi\'c et al. \cite{Ivezic2002}). 
A high amount of radio-loud quasars was also found for the WLQ sample from Paper 1. In addition, we constructed
a sample of 98 WLQs from the literature (Diamond-Stanic et al. \cite{Diamond2009};
Shemmer et al. \cite{Shemmer2010}; Lane et al. \cite{Lane2011};  Niko{\l}ajuk \& Walter \cite{Nikolajuk2012};
Wu et al. \cite{Wu2012}). The percentage of radio-loud quasars is lower for that sample.
However, the majority of these quasars have higher redshifts than ours and when
we reject those with  $z>3$, the remaining subsample has a radio-loudness rate as
high as in our WLQ sample. The lower radio-loudness percentage among the high redshift WLQs from the 
literature is mainly due to the fact that many of these sources were selected to be not radio-loud.
Finally, we selected all quasars with \ion{Mg}{ii} equivalent widths
$W_{\ion{Mg}{ii}}<15$\AA\ from Shen et al. (\cite{Shen2011}). Again, the radio-loudness proportion
(0.22) is much higher than typical for the parent quasar sample. On the other hand, though $f_{\rm rl}$ 
is high,  the mean radio-loudness $\overline{R}$ for the WLQs is much lower than for
the comparison sample.

\begin{figure}[ht]
\begin{centering}
\includegraphics[viewport=60 20 610 785,angle=270,width=8.8cm,clip]{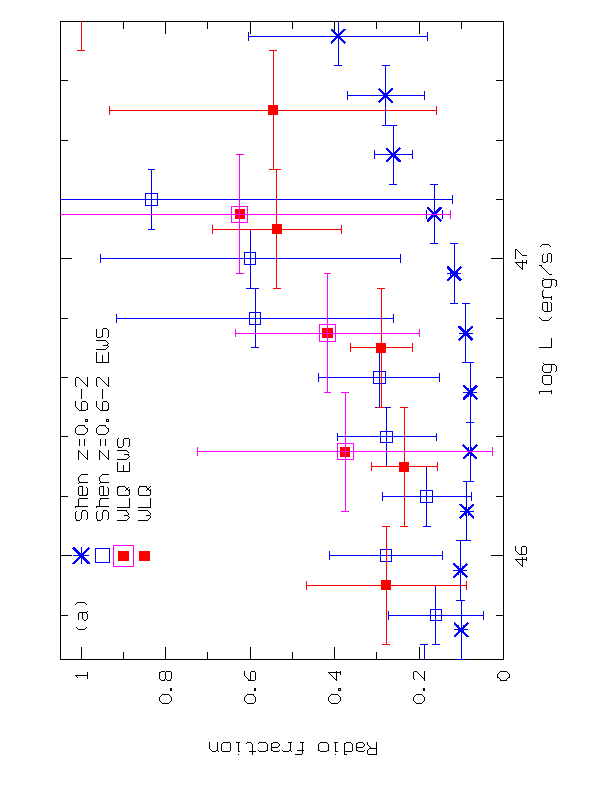} 
\includegraphics[viewport=60 20 610 785,angle=270,width=8.8cm,clip]{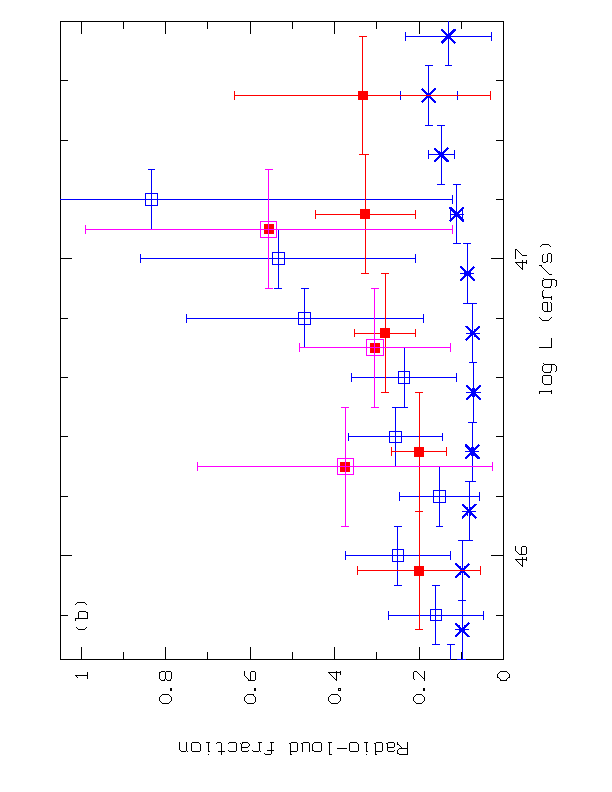}
\end{centering}
\caption{Fraction of (a) FIRST radio-detected quasars and (b) radio-loud quasars in luminosity bins for
the quasars from the Shen catalogue (blue asterisks), the subsample of Shen quasars with 
$W_\ion{Mg}{ii}<11$\,\AA\ and $W_\ion{C}{iv}<4.8$\,\AA\ (open blue squares), our WLQ sample 
(filled red squares), and the WLQ-EWS subsampls of WLQs (magenta framed squares).
All samples restricted to $0.6\le z \le 2$.
}
\label{fig:radio_fraction}
\end{figure}

\begin{figure*}[ht]
\begin{centering}
\includegraphics[viewport=0 40 600 785,angle=0,width=9.0cm,clip]{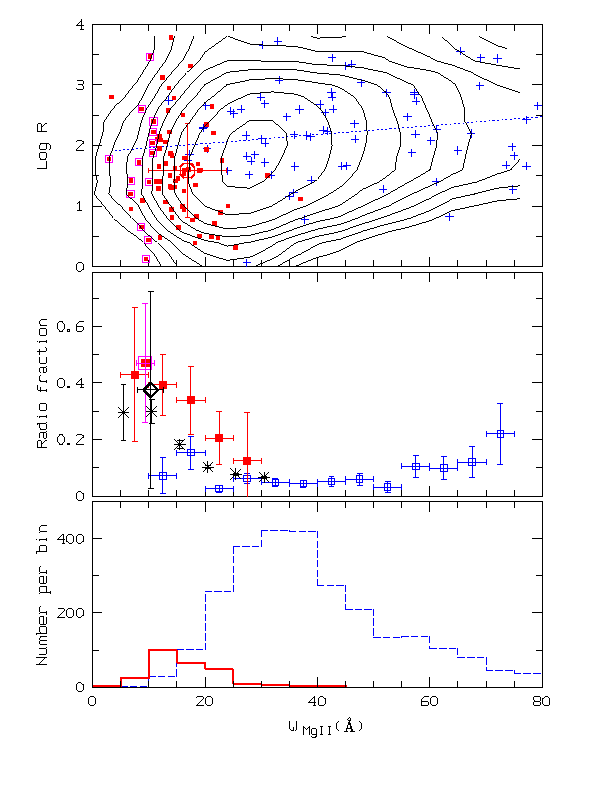}
\includegraphics[viewport=0 40 600 785,angle=0,width=9.0cm,clip]{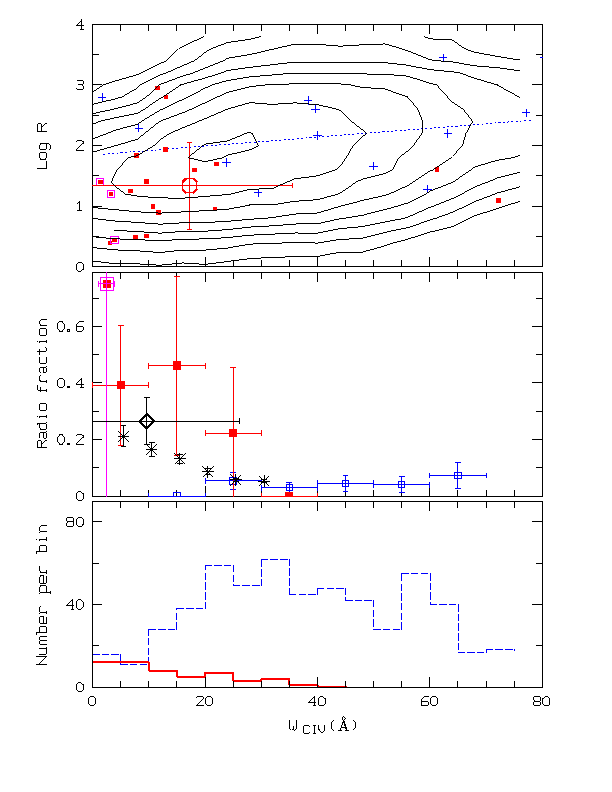}
\end{centering}
\caption{Radio loudness and equivalent width of \ion{Mg}{ii} (left) and \ion{C}{iv} (right). 
Top: radio-loudness parameter $R$ 
(filled red squares: rWLQ sample, 
magenta framed filled squares: rWLQ-EWS sample,
large open red circle with error bars: mean value and 1$\sigma$ error of rWLQs,
blue plus signs: comparison sample, contours: all quasars from the Shen catalogue with {\sc bal\_flag = 0}). 
Middle: proportion $f_{\rm r}$ of FIRST-detected quasars in EW bins
(filled red squares: rWLQ sample, 
magenta framed filled square: mean value from the rWLQ-EWS sample,
open blue squares: comparison sample,  
black lozenge: WLQ Lit ($z<3$) sample, 
black asterisks: all quasars from the Shen catalogue with {\sc bal\_flag = 0}, 
horizontal bars: binning intervals,  
vertical bars: uncertainties from Poisson statistics and error propagation).
Bottom: histogram distribution of the EW 
(solid red: rWLQ sample, dashed blue: comparison sample).
}
\label{fig:EW_radio}
\end{figure*}

The proportion $f_{\rm r} \equiv N_{\rm r; R>0}/N_{\rm all}$ of the radio-detected 
quasars in the FIRST survey area amounts to 0.34 for the entire WLQ sample,
0.33 for the rWLQ sample, but only 0.06 for the comparison sample.
Again, the ratios are higher for the EWS subsamples (0.42 and 0.47).
The radio proportion is also higher for the other WLQ samples 
in Tab.\,\ref{tab:radio-prop}. It is qualitatively expected that the percentage of radio detections 
is higher for our WLQ sample than for the comparison sample because of the higher luminosity. 
As a consequence of the relatively narrow $z$ range, the luminosity strongly correlates with the flux 
density and  the WLQ sample is thus expected to also have a higher number of quasars with 
1.4 GHz radio flux densities exceeding the FIRST detection limit, even if $R$ is independent of $L$.

Figure\,\ref{fig:radio_fraction}  shows the proportion of (a) the radio-detected quasars 
and (b) the radio-loud quasars as functions of the luminosity
for both WLQs and ordinary quasars from the Shen catalogue. 
Both samples were restricted to $z = 0.6-2.0$, 
where the overwhelming majority (93\%) of the WLQs are found. For the Shen catalogue sample,
the trend with  $L$ is only moderate for the proportion of radio-loud quasars but strong 
for the proportion of radio detections. On the other hand, 
Fig.\,\ref{fig:radio_fraction} clearly shows that the luminosity dependence of the 
proportion of radio detections is by far not strong enough to explain the high rates  
of FIRST-detected WLQs. 
The WLQ-EWS subsample follows the same trend as our visually selected WLQ sample. 
We conclude that the high percentage of radio sources among our WLQs is not primarily a 
consequence of the higher luminosities, i.e. of the Baldwin effect in combination with 
the S/N selection bias discussed in Sect.\,\ref{subsec:lum_etc}.

In the top panels of Fig.\,\ref{fig:EW_radio}, we plotted $R$ as a function of
$W_{\rm \ion{Mg}{ii}}$ (left) and $W_{\rm \ion{C}{iv}}$ (right), respectively,
for the rWLQ sample and the comparison sample of normal quasars. In addition, the distributions
for all quasars from the Shen catalogue is shown by contour lines.
There is a wide scatter, but the linear regressions for the comparison sample indicates
an increase of $R$ with increasing $W$. The centroid of the WLQ distribution is slightly
below the regression line. The middle panels of Fig.\,\ref{fig:EW_radio} show
the proportins $f_{\rm r}(W)$ of radio-detected quasars in EW bins. There seems to be a
negative correlation between  $f_{\rm r}$ and $W$ for the WLQs.

The overwhelming majority of the FIRST-detected WLQs are core-dominant radio sources. 
The ratio $r_{N, {\rm cl}} = N_{\rm c}/N_{\rm l}$ of the numbers of core-dominant to 
lobe-dominant sources is about three times higher than for the 
comparison sample. Again, high ratios $r_{N, {\rm cl}}$ are also found for the other WLQ samples. 
On the other hand, the ratio $r_{R, {\rm cl}} = \overline{R}_{\rm c}/\overline{R}_{\rm l}$
of the mean radio-loudness of the core-dominant 
to that of the lobe-dominant sources is about $2\ldots 3$ times lower than usual.
The very low value of $r_{R, {\rm cl}} = 0.04$ for the rWLQ-EWS subsample is likely an effect of the 
poor statistics with only two lobe-dominated quasars.

\subsection{X-rays}

Finally, we briefly consider the rate of X-ray sources. We use the column 
{\sc rass\_offset} from the Shen catalogue to select X-ray detections. For an offset less
then 30 arcsec, we find a percentage of 6.5\% for the rWLQ sample compared to
2.6\% for the comparison sample. If we restrict the offset to less than 10 arcsec, the
corresponding proportions are 1.8\% and 0.6\%, respectively.   Therewith, the X-ray percentage 
appears  to be three times higher for the WLQs than for the ordinary quasars,
but this low-number statistics is very uncertain. The mean redshifts and RASS 
count rates  are similar for both samples.

%
%
\section{Discussion}\label{sec:discussion}
%
%

We found that the continuum of the composite SED for the WLQ sample is steeper than for ordinary quasars. The 
WLQ composite spectrum is reasonably well matched by a modified composite of normal quasars where the BELs are normal but 
the continuum is hotter than usual. We also found that the WLQs have higher luminosities, Eddington ratios, 
and accretion rates, while the variability is lower. Finally, the WLQ sample is also significantly different from 
the comparison sample of ordinary quasars with respect to the radio properties. In this section, we try to 
give a consistent interpretation of these findings.

In the standard picture (Shakura \& Sunyaev \cite{Shakura73}; Frank et al. \cite{Frank02}), the 
temperature of the accretion disk around a black hole of given mass is determined by the 
accretion rate. Higher accretion rates lead to higher disk temperatures, luminosities, and Eddington ratios. 
The variability is known to anti-correlate with the Eddington ratio and the accretion rate
(Wilhite et al. \cite{Wilhite2008}; Bauer et al. \cite{Bauer2009}; Zuo et al. \cite{Zuo2012}; 
Meusinger \& Weiss \cite{Meusinger2013}). 
The accretion process in quasars is probably accompanied by strong local temperature fluctuations
(Dexter \& Agol \cite{Dexter2011}; Dexter et al. \cite{Dexter2012}; Ruan et al. \cite{Ruan2014}) and 
perhaps also by global fluctuations of the accretion rate (Pereyra et al. \cite{Pereyra2006}; 
Li \& Cao \cite{Li2008}; Zuo et al. \cite{Zuo2012}). The variability strength of the 
BEL flux is an order of magnitude less than for the underlying continuum  (Wilhite et al. \cite{Wilhite2005}; 
Meusinger et al. \cite{Meusinger2011}). A change of the level of the continuum flux is not 
immediately accompanied by a change of the line flux on the same level. 
Hryniewicz et al. (\cite{Hryniewicz2010}) proposed a scenario where quasar activity has an 
intermittent character with several subphases. Each subphase starts with a slow development of the BLR.
At least in a statistical sense, the WLQs in our sample can be consistently understood as AGNs
in the beginning of a phase of stronger accretion, i.e. accretion rate and luminosity enhanced and
variability thus reduced, whereas the BLR has not
yet adapted to the level of the disk. In such a scenario, high ionisation BELs, like
\ion{C}{iv} and Ly$\alpha$, are expected to be weaker than low ionisation BELs, 
like \ion{Mg}{ii}, because they are formed at a higher distance from the disk. In fact, 
our WLQ sample indicates more weakening of the high ionisation compared to the low 
ionisation lines (Tab.\,\ref{tab:mean-prop}; Fig.\,\ref{fig:fit}).

In order to account for the high amount of radio-detected quasars among our WLQs, we consider 
the viewing angle towards a radio jet as another possibly important aspect for the WLQ phenomenon. 
We found that our WLQs are, on average, less radio-loud than ordinary quasars but are much more
likely to have FIRST counterparts. About one third of the WLQs in our sample are radio sources 
on the FIRST level.

First, however, we have to make sure that the high percentage of radio detections is not a selection effect.
In Paper\,1, we found remarkably high radio detection rates for different types of unusual quasars, 
particularly for unusual BAL quasars. The radio detection rate was found there to positively 
correlate with the degree of the deviation from the SDSS quasar composite spectrum. 
The ratio, $N_{\rm F}/N_{\rm C}$,  of the number of quasars with the FIRST target flag set to the number
of quasars with their colour target flag set was found there significantly higher for the unusual quasars
(0.33) than for the entire quasar catalogue (0.07). In addition, we considered the ratio 
$N_{\rm Fs}/N_{\rm Cs}$, where $N_{\rm Fs}$ is the number of quasars selected solely by the FIRST 
selection but not the colour selection and $N_{\rm Cs}$ is the number of quasars without FIRST 
counterparts but selected solely based on their colours. Again, we found a much higher value (0.22) 
for the unusual than for the entire quasar sample (0.01).
For the WLQ sample of the present study, we have $N_{\rm F}/N_{\rm C} = 0.33$, reflecting once again 
the high proportion of radio detections, but $N_{\rm Fs}/N_{\rm Cs} = 0.01$, which is in accordance with 
the ordinary quasars. The latter is actually not surprising because the SDSS 
colour-$z$ relations of the WLQs are not much different from the mean relations (Fig.\ref{fig:SDSS_WISE}).
This indicates that, different from e.g. the unusual BAL quasars in Paper 1, the high radio detection rate 
of the WLQs cannot be unambiguously attributed to a selection bias in the sense that a high number
were only targeted by SDSS because they had been detected as radio sources. As a consequence, we 
have to assume that the high percentage of radio sources is an intrinsic property.

Our WLQ sample was visually selected from the Kohonen icon maps (Sect.\,\ref{sec:selection}).
The basic criterion for the inclusion into the final sample was the presence of weak but clearly
identifiable BELs. Moreover, we only accepted objects identified in the SDSS DR7 quasar catalogue 
(Schneider et al. \cite{Schneider2010}; Shen et al. \cite{Shen2011}) where the presence of broad 
lines in the spectra is required to be included. The presence of BELs is usually considered as
evidence that the continuum radiation is not dominated by a beamed synchrotron component.
Therefore our sample is unlikely to be dominated by BL Lac objects where the beamed synchrotron emission
strongly overpowers the thermal emission. The variability data (Sect.\,\ref{subsec:variability}) are also 
inconsistent with a dominance of beamed non-thermal emission.

It has been suggested that the radio-to-optical flux ratio $R_{\rm c}$ from the radio core is a useful 
statistical measure of orientation (Baker \& Hunstead \cite{Baker1995}; Wills \& Brotherton \cite{Wills95}; 
Kimball et al. \cite{Kimball11}). In standard unification theory, orientation has an effect on 
the optical spectrum both via obscuration in the plane of the accretion flow and via relativistic 
boosting in the line of sight close to the relativistic outflow. High values of $R_{\rm c}$ are 
supposed to indicate a low angle to the jet axis. We did not explicitly separate the radio 
emission into core and extended components. However, because the majority ($>90$\%) of our radio-detected 
WLQs are core dominant, we simply identify $R$ with $R_{\rm c}$.  Then, Fig.\,\ref{fig:EW_radio} 
would mean that our radio-detected WLQs have higher inclination angles to the jet. A similar trend of $R_{\rm c}$ with 
$W_\ion{Mg}{ii}$ and $W_\ion{C}{IV}$ was reported by Kimball et al. (\cite{Kimball11}) who predicted that
this correlation may be caused by anisotropic emission from the BLR. From the study of composite spectra of 
quasar samples grouped by the radio core-to-lobe ratio, Baker \& Hunstead (\cite{Baker1995}) found 
stronger reddening in lobe-dominated quasars thought to be observed at higher inclination angles to 
the jet (i.e. edge-on). Because the core-to-lobe ratio is correlated with $R_{\rm c}$ 
(Kimball et al. \cite{Kimball11}), we would expect that our WLQs exhibit stronger reddened spectra 
because of their lower mean $R$ value. This, however, is clearly not the case. The median SED of the WLQs is 
steeper than that of the comparison sample. If this would be caused by reddening, the WLQ sample would be 
less reddened. We conclude that the WLQs in our sample are not seen preferentially edge on.

The majority (65\%) of the WLQs with $R>10$ have radio-loudness parameters in the range $25\ldots 250$, sometimes
called radio-intermediate. It was proposed that flat-spectrum radio-intermediate quasars are 
relativistically boosted radio-quiet quasars (Falcke et al. \cite{Falcke1996}; Wang et al. \cite{Wang2006}). 
If this also applies to the radio-intermediate WLQs in our sample, we expect to view these quasars
at low inclination angles to the jet.
We argued (Fig.\,\ref{fig:wide_band}) that the difference between the slopes of the median wide band 
SEDs of the WLQ sample and the comparison sample can be explained by an additional power-law component,
respectively a hotter continuum (Fig.\,\ref{fig:fit}), that does not dominate but makes a substantial 
contribution to the observed flux of the WLQs.  An enhancement of the continuum by a factor of two caused by 
an additional component reduces the EW of the lines correspondingly (Fig.\,\ref{fig:fit}).

%
%
\section{Summary and conclusions}\label{sec:summary}
%
%

We performed a new search for quasars with weak emission lines in the spectroscopic data from the 
SDSS DR7. We visually inspected the 36 self-organising maps (Kohonen maps)  from Paper 1 for nearly $10^5$ spectra 
classified as quasars by the SDSS spectroscopic pipeline and selected a sample of $\sim 2500$ WLQ candidates. After the
thorough individual analysis of all selected spectra we created a final sample of 365 WLQs with mean redshift
$z = 1.50 \pm 0.45$. The mean equivalent widths are $17 \pm 8$ \AA\ for \ion{Mg}{ii} and $13 \pm 14$ \AA\ for 
\ion{C}{iv}. 
To avoid contamination, featureless spectra were not included. The corresponding incompleteness is estimated to 
about 10\%. The sample includes a subsample of 46 WLQs with EWs below 3-sigma thresholds defined by the EW 
distribution of the ordinary quasars (EWS subsample). Especially for the analysis of the accretion rates, 
a subsample restricted to the redshift range $0.7 < z < 1.7$ was considered (rWLQ sample).

The investigation of the properties of the WLQs (WLQ-EWS subsample, rWLQ-EWS subsample, and their parent 
samples) and their comparison with corresponding control samples yields the following results:

\begin{itemize}
 \item 
 The SDSS composite spectrum for the WLQs shows significantly weaker BELs and a bluer continuum compared 
 to the ordinary quasars. Therefore it can be excluded that WLQs are stronger affected by dust reddening
 than usual and it is unlikely that WLQs are
 preferentially seen edge-on. No significant differences are found between the composites of the radio-loud and 
 the not radio-loud WLQs. (Sect. \ref{subsec:composite})
 \item
 The wide-band SED constructed from the SDSS, 2MASS, and WISE  photometric data
 (Sect. \ref{subsec:wide-band}) shows that the trend of a steeper continuum continues towards the mid infrared. 
 \item
 The WLQs have, on average, significantly higher luminosities, Eddington ratios, and accretion rates,
 but not significantly higher black hole masses (Sect. \ref{subsec:lum_etc}). 
 About half of the luminosity excess is produced by a S/N bias in the selection process.
 The remaining intrinsic luminosity excess corresponds to the Baldwin effect and also manifests by
 higher accretion rates and Eddington ratios. The higher mean mass is simply a consequence of the
 higher mean luminosity in combination with the positive $M - L$ correlation in the parent sample.
 \item
 Indicators for the strength of the UV and optical variability are available for 23 WLQs from our sample. 
 They show a wide scatter with a tendency towards lower variability than
 that of the ordinary quasars of comparable luminosities
 (Sect. \ref{subsec:variability}).
 \item
 The WLQ sample has remarkable radio properties (Sect. \ref{subsec:radio}) that are probably not produced by
 selection effects. The percentage of radio-detected quasars is more than five times higher then for the 
 control sample and the ratio of the number of core-dominant to the number of lobe-dominant radio sources 
 is about three times higher. On the other hand, the mean radio-loudness of the radio-detected WLQs is 
 much lower than for the ordinary quasars. The propotion of radio-sources increases towards lower equivalent 
 widths while the mean radio loudness decreases.
\end{itemize}

The higher luminosities and Eddington ratios in combination with a bluer SED can be consistently explained by hotter
accretion disks, i.e. by stronger accretion. According to the scenario proposed by Hryniewicz et al. (\cite{Hryniewicz2010}),
a change towards a higher accretion rate is accompanied by an only slow development of the BLR. We have shown 
that the composite WLQ spectrum can be reasonably matched by the ordinary quasar composite where the 
continuum has been replaced by that of a hotter disk (Fig. \ref{fig:fit}). Therewith, at least 
a substantial percentage of the WLQs can be normal quasars in an early stage of increased accretion activity. 
On the other hand, a similar effect can be achieved by an additional power-law component, perhaps in relativistically
boosted radio-quiet quasars viewed at low inclination angles to the jet.

Our WLQ sample is thus probably a mixture of quasars at the beginning of a stage of increased accretion
activity on the one hand side and of beamed radio-quiet quasars on the other. There are hints on a close 
link between the accretion process and the relativistic jets (e.g.  Falcke \& Biermann \cite{Falcke1995};
Rawlings \& Saunders \cite{Rawlings1991}; Cao \& Jiang \cite{Cao1999}; Dexter et al. \cite{Dexter2012}) 
and one may expect that the two scenarios are closely linked to each other where an
initially high or an enhanced accretion rate is the reason for the relatively high rate of
radio detected quasars.

%
\begin{acknowledgements} 
We thank the anonymous referee for suggestions that significantly improved our manuscript.
We further thank Martin Haas for valuable comments and tips.
This research has made use of data products from the Sloan Digital Sky Survey (SDSS),
the Two Micron All-Sky Survey (2MASS), and the Wide-Field Infrared Survey (WISE).
Funding for the SDSS and SDSS-II has been provided by the Alfred P. Sloan Foundation,
the Participating Institutions (see below), the National Science Foundation, the National
Aeronautics and Space Administration, the U.S. Department of Energy, the Japanese
Monbukagakusho, the Max Planck Society, and the Higher Education Funding Council for
England. The SDSS Web site is http://www.sdss.org/. The SDSS is managed by the Astrophysical Research
Consortium (ARC) for the Participating Institutions. The Participating Institutions are: the American
Museum of Natural History, Astrophysical Institute Potsdam, University of Basel, University of Cambridge
(Cambridge University), Case Western Reserve University, the University of Chicago, the Fermi National
Accelerator Laboratory (Fermilab), the Institute for Advanced Study, the Japan Participation Group,
the Johns Hopkins University, the Joint Institute for Nuclear Astrophysics, the Kavli Institute for
Particle Astrophysics and Cosmology, the Korean Scientist Group, the Los Alamos National Laboratory,
the Max-Planck-Institute for Astronomy (MPIA), the Max-Planck-Institute for Astrophysics (MPA),
the New Mexico State University, the Ohio State University, the University of Pittsburgh, University
of Portsmouth, Princeton University, the United States Naval Observatory, and the University of
Washington. 
The Two Micron All Sky Survey is a joint project of the University of Massachusetts
and the Infrared Processing and Analysis Center/California Institute of Technology,
funded by the National Aeronautics and Space Administration and the National Science
Foundation.
The Wide-field Infrared Survey Explorer is a joint project of the University of California, Los Angeles,
and the Jet Propulsion Laboratory/California Institute of Technology, funded by the National
Aeronautics and Space Administration.
\end{acknowledgements}
%
%
%
{}


\begin{thebibliography}{}
\bibitem[2009]{Abazajian2009}
  Abazajian, K. N., Adelman-McCarthy, J. K., Ag\"ueros, M. A., et al.\ 2009, \apjs, 182, 543
\bibitem[2010]{Ai2010}
  Ai, Y., Yuan, W., Zhou, H., et al.\ 2010, \apj, 716, 31
\bibitem[1995]{Baker1995}
  Baker, J. C., \& Hunstead, R. W.\ 1995, \apj, 286, 23
\bibitem[2004]{Bachev2004}  
  Bachev, R., Marziani, P., Sulentic, J. W., et al.\ 2004, \apj, 617, 171
\bibitem[1977]{Baldwin1977}
  Baldwin, J.\ 1977, \apj, 214, 679
\bibitem[2004]{Baskin2004}
  Baskin, A. \& Laor, A.\ 2004, MNRAS, 350, L31  
\bibitem[2009]{Bauer2009}
  Bauer, A., Baltey, C., Coppi, P., et al. \ 2009, \apj, 705, 46
\bibitem[1995]{Becker1995}
  Becker, R. H., White, R. L., \& Helfand, D. J. \ 1995, \apj, 450, 559 
\bibitem[2012]{Bian2012}
  Bian, W.-H., Fang, L.-L., Huang, K.-L., et al.\ 2012, \mnras, 427, 2881  
\bibitem[2008]{Bramich08}
  Bramich, D. M., Vidrih, S., Wyrzykowski, L., et al. \ 2008, \mnras, 386, 887 
\bibitem[1999]{Cao1999}
  Cao, X. \& Jiang, D. R.\ 1999, \mnras, 307, 802 
\bibitem[2002]{Comastri2002}
  Comastri, A., Mignoli, M., Ciliegi, P., et al.\ 2002, \apj, 571, 771
\bibitem[2011]{Davis2011}
  Davis, S. W. \& Laor, A. \ 2011, \apj, 728, 98
\bibitem[2005]{DeVries2005}
  De Vries, W. H., Becker, R. H., White, R. L., \& Loomis, C. \ 2005, \aj, 129, 615
\bibitem[2011]{Dexter2011}
  Dexter, S. W. \& Agol, E. \ 2011, \apj, 727, L24
\bibitem[2012]{Dexter2012}
  Dexter, J., McKinney, J. C., \& Agol, E. \ 2012, \mnras, 421, 1517 
\bibitem[2009]{Diamond2009}
  Diamond-Stanic, A. M., Fan, X., Brandt, W., et al.\ 2009, \apj, 699, 782
\bibitem[2002]{Dietrich2002}
  Dietrich, M., Hamann, F., Shields, J. C., et al.\ 2002, \apj, 581, 912
\bibitem[2009]{Dong2009}
  Dong, X.-B., Wang, T.-G., Wang, J.-G., et al.\ 2009, \apj, 703, L1
\bibitem[2011]{Dong2011}
  Dong, X.-B., Wang, J.-G., Ho, L. C., et al.\ 2011, \apj, 736, 86   
\bibitem[1995]{Falcke1995}
  Falcke, H. \& Biermann, P. \ 1995, A\&A, 293, 665
\bibitem[1996]{Falcke1996}
  Falcke, H., Sherwood, W., \& Patnaik, A. R.\ 1996, \apj, 471, 106
\bibitem[1999]{Fan1999}
  Fan, X., Strauss, M. A., Gunn, J. E., et al.\ 1999, \apj, 526, L57
\bibitem[1993]{Francis1993}
  Francis, P. J., Hooper, E. J., \&  Impey, C. D. \ 1993, \aj, 106, 417 
\bibitem[2002]{Frank02}
  Frank, J., King, A. R.,\&  Raine, D. J. \ 2002, Accretion Power in Astrophysics (third edition),
  Cambridge University Press 
\bibitem[2011]{Ghisellini2011}
  Ghisellini, G., Tavecchio, F., Foschini, L., et al.\ 2012, \mnras, 414, 2674
\bibitem[2011]{Girven2011}
  Girven, J., G\"ansicke, B. T., Steeghs, D., \& Koester, D.\ 2011, \mnras, 417, 1210
\bibitem[2001]{Green2001}
  Green, P. J., Forster, K., \& Kuraszkiewicz, J.\ 2001, \apj2, 556, 727  
\bibitem[1994]{Hook1994}
  Hook, I. M., McMahon, R. G., Boyle, B. J., \& Irwin, M. J. \ 1994, \mnras, 268, 305
\bibitem[2010]{Hryniewicz2010}
  Hryniewicz, K., Czerny, B., Niko{\l}ajuk, M., \& Kuraszkiewicz, J.\ 2010, \mnras, 404, 2028
\bibitem[2012]{inderAu2012}
  in der Au, A., Meusinger, H., Schalldach, P., \& Newholm, M. \ 2012, A\&A, 547, A115
\bibitem[2002]{Ivezic2002}
  Ivezi\'c, \v{Z}., Menou, K., Knapp, G. R., et al. \ 2002, \aj, 124, 2364
\bibitem[2007]{Jiang2007}
  Jiang, L., Fan, X., Ivezi\'c, \v{Z}., et al. \ 2007, \apj, 656, 680
\bibitem[1989]{Kellermann1989}
  Kellermann, K. I., Sramek, R., Schmidt, M., et al. \ 1989, \aj, 98, 1195
\bibitem[2011]{Kimball11}
  Kimball, A. E., Ivezi\'c, \v{Z}., Wiita, P., J., \& Schneider, D. P., 2011, \aj, 141, 182
\bibitem[2013]{Kleinman2013}
  Kleinman, S. J., Kepler, S. O., Koester, D., et al. \ 2013, \apjs, 204, 5  
\bibitem[2001]{Kohonen2001}
  Kohonen, T. \ 2001, Self-Organizing Maps, third edition, New York: Springer
\bibitem[2011]{Lane2011}
  Lane, R. A., Shemmer, O., Diamond-Stanic, A. M., et al.\ 2011, \apj, 743, 163
\bibitem[2011]{Laor2011}
  Laor, A. \& Davis, S. W.\ 2010, \mnras, 417, 681-688
\bibitem[1988]{Lawrence1988}
  Lawrence, A., Saunders, W., Rowan-Robinson, M., et al.\ 1988, \mnras, 235, 261
\bibitem[2007]{Leighly2007}
  Leighly, K. M., Halpern, J. P., Jenkins, E. B., et al.\ 2007, \apjs, 173, 1
\bibitem[2008]{Li2008}
  Li, S.-L. \& Cao, X. \ 2008, \mnras, 387, L41
\bibitem[2009]{Massaro2009}
  Massaro, E., Giommi, P., Leto, C., et al. \ 2009, \aap, 495, 691
\bibitem[1999]{McCook1999}
  McCook, G. P. \& Sion, E. M. \ 1999, \apjs, 121, 1
\bibitem[1995]{Mcdowell1995}
  McDowell, J. C., Canizares, C., Elvis, M., et al.\ 1995, \apj, 450, 585
\bibitem[2004]{McLure2004}
  McLure, R. J. \& Jarvis, M. J.\ 2004, \mnras, 353, L45
\bibitem[2011]{Meusinger2011}
  Meusinger, H., Hinze, A., \& de Hoon, A.\ 2011, A\&A, 525, A37
\bibitem[2012]{Meusinger2012}
  Meusinger, H., Schalldach, P., Scholz, R., et al.\ 2012, A\&A, 541, A77 (Paper 1)
\bibitem[2013]{Meusinger2013}
  Meusinger, H. \& Weiss, V.\ 2013, A\&A, 560, A104
\bibitem[1992]{Netzer1992}
  Netzer, H., Laor, A., \& Gondhalekar, P. M. \ 1992, \mnras, 254, 15
\bibitem[2012]{Nikolajuk2012}
  Niko{\l}ajuk, M. \& Walter, R.\ 2012, \mnras, 420, 2518
\bibitem[1997]{Paltani1997}
  Paltani, S. \& Courvoisier, T. \ 1997, A\&A, 291, 74
\bibitem[2006]{Pereyra2006}
  Pereyra, N. A., Vanden Berk, D. E., Turnshek, D. A., et al. \ 2006, \apj, 642, 87
\bibitem[1983]{Pica1983}
  Pica, A. J. \& Smith, A. G. \ 1983, \apj, 272, 11
\bibitem[2008]{Plotkin2008}
  Plotkin, R. M., Anderson, S. F., Hall, P. B., et al.\ 2008, \aj, 135, 2453  
\bibitem[2010]{Plotkin2010}
  Plotkin, R. M., Anderson, S. F., Brandt, W., et al.\ 2010, \apj, 721, 562
\bibitem[1991]{Rawlings1991}
  Rawlings, S. G. \& Saunders, R. D. \ 1991, Nature, 349, 138
\bibitem[2001]{Richards2001}
  Richards, G. T., Fan, X., Schneider, D. P., et al. \ 2001, \aj, 121, 2308
\bibitem[2011]{Richards2011}
  Richards, G. T., Kruczek, N. E., Gallagher, S. C., et al. \ 2011, \aj, 141, 167
\bibitem[2010]{Roeser2010}
  R\"oser, S., Demleitner, M., \& Schilbach, E. \ 2010, \aj, 139, 2440
\bibitem[2014]{Ruan2014}
  Ruan, J. J., Anderson, S. F., Dexter, E., \& Agol, E. \ 2014, arXiv:1401.1211v1
\bibitem[2010]{Schneider2010}
  Schneider, D. P., Richards, G. T., Hall, P. B., et al.\ 2010, \aj, 139, 2360
\bibitem[2003]{Severgnini2003}
  Severgnini, P., Caccianiga, A., Braito, V., et al.\ 2003, A\&A, 406, 483
\bibitem[1973]{Shakura73}
  Shakura, N. I. \& Sunyaev, R. A. \ 1973, A\&A, 24, 337 
\bibitem[2003]{Shang2003}
  Shang, Z., Wills, B. J., Robinson, E. L., et al.\ 2003, \apj, 586, 52
\bibitem[2006]{Shemmer2006}
  Shemmer, O., Brandt, W., Netzer, H., et al.\ 2006, \apj, 646, L29
\bibitem[2010]{Shemmer2010}
  Shemmer, O., Trakhtenbrot, B., Anderson, S. F., et al.\ 2010, \apj, 722, L152
\bibitem[2011]{Shen2011}
  Shen, Y., Richards, G. T., Strauss, M. A., et al.\ 2011, \apjs, 194, 45
\bibitem[2012]{Shen2012}
  Shen, Y. \& Liu, X. \ 2012, \apj, 753, 125  
\bibitem[2010]{Shi2010}
  Shi, Y., Rieke, G. H., Smith, P., et al.\ 2010, \apj, 714, 115
\bibitem[1988]{Siegel1988}
  Siegel, S. \& Castellan, N. J. \ 1988, Nonparametric Statistics for the Behavioral Sciences, 
  (New York: McGraw-Hill, Inc.)
\bibitem[2006]{Skrutskie2006}
  Skrutskie, M. F., Cutri, R. M., Stiening, R., et al.\ 2006, \aj, 131, 1163
\bibitem[2012]{Thorstensen2012}
  Thostensen, J. R. \& Skinner, J. N.\  2012, \aj, 144, 81 
\bibitem[2003]{Tran2003}
  Tran, H. D.\ 2003, \apj, 583, 632
\bibitem[1995]{Urry1995}
  Urry, C. M. \& Padovani, P. \ 1995, \pasp, 107, 803
\bibitem[2001]{VandenBerk2001}
  Vanden Berk, D. E., Richards, G. T., Bauer, A., et al. \ 2001, \aj, 122, 549
\bibitem[2004]{VandenBerk2004}
  Vanden Berk, D. E., Wilhite, B. C., Kron, R. G., et al. \ 2004, \apj, 601, 692
\bibitem[2006]{Wang2006}
  Wang, T.-G., Zhou, H.-Y., Wang, J.-X., et al. \ 2006, \apj, 645, 856
\bibitem[2000]{White2000}
  White, R. L., Becker, R. H., Gregg, M. D., et al. \ 2000, \apjs, 126, 133
\bibitem[2005]{Wilhite2005}  
  Wilhite, B. C., Vanden Berk, D. E., Kron, R. G., et al. \ 2005, \apj, 633, 638
\bibitem[2008]{Wilhite2008}  
  Wilhite, B. C., Brunner, R. J., Grier, C. J., et al. \ 2008, \mnras, 383, 1232
\bibitem[1995]{Wills95}
  Wills, B. J. \& Brotherton, M. S. \ 1995, \apj, 448, L81 
\bibitem[2010]{Wright2010}
  Wright, E. L., Eisenhardt, P. R. M., Mainzer, A. K., et al.\ 2010, \aj, 140, 1868
\bibitem[2009]{Wu2009}
  Wu, J., VandenBerk, D. E., Brandt, W. N., et al.\ 2009, \apj, 702, 767
\bibitem[2011]{Wu2011}
  Wu, J., Brandt, W., Hall, P. B., et al.\ 2011, \apj, 736, 28
\bibitem[2012]{Wu2012}
  Wu, J., Brandt, W., Anderson, S. F., et al.\ 2012, \apj, 747, 10
\bibitem[2000]{York2000}
  York, D. G., Adelman, J., Anderson, J. E., Jr., et al. \ 2000, \aj, 120, 1579
\bibitem[1993]{Zheng1993}
  Zheng, W. \& Malkan, M. A. \ 1993, \apj, 415, 517 
\bibitem[2012]{Zuo2012}
  Zuo, W., Wu, X.-B., Liu, Y.-Q., \& Jiao, C.-L. \ 2012, \apj, 758, 104 
\end{thebibliography}
\end{document}